\documentclass[manuscript]{aastex}

\newcommand{\teff}{\ensuremath{T_{\rm eff}}}

\shorttitle{Stellar Classification for Kepler}
\shortauthors{T.M. Brown, D.W. Latham, M. Everett}

{\obeyspaces\gdef {\leavevmode\space}}

\begin{document}
\title{Kepler Input Catalog: Photometric Calibration and Stellar Classification}
\author{Timothy M. Brown}
\affil{Las Cumbres Observatory Global Telescope, Goleta, CA 93117}
\email{tbrown@lcogt.net}
\author{David W. Latham}
\affil{Harvard-Smithsonian Center for Astrophysics, Cambridge, MA 02138}
\email{latham@cfa.harvard.edu}
\author{Mark E. Everett}
\affil{National Optical Astronomy Observatories, Tucson, AZ 85721}
\email{everett@noao.edu}
\author{Gilbert A. Esquerdo}
\affil{Harvard-Smithsonian Center for Astrophysics, Cambridge, MA 02138}
\email{gesquerd@cfa.harvard.edu}

\begin{abstract}

We describe the photometric calibration and stellar classification methods
used by the Stellar Classification Project (SCP)
to produce the
{\it Kepler} Input Catalog (KIC).
The KIC is a catalog containing photometric and physical data for sources
in the {\it Kepler Mission} field of view;
it is used by the mission to select optimal targets.
Four of the visible-light ($g,r,i,z$)  magnitudes used in the KIC are 
tied to SDSS magnitudes; the fifth ($D51$) is an AB magnitude calibrated
to be consistent with \citet{cas04} (henceforth CK) model atmosphere fluxes.
We derived atmospheric extinction corrections from hourly observations
of secondary standard fields within the {\it Kepler} field of view.
For these filters and extinction estimates, repeatability of 
absolute photometry for stars
brighter than magnitude 15 is typically 2\%. 
We estimated stellar parameters \{$\teff$, $\log(g)$, $\log (Z)$,
$E_{B-V}$\} using Bayesian posterior probability maximization 
to match observed
colors to CK stellar atmosphere models.
We applied Bayesian priors describing the distribution of solar-neighborhood
stars in the color-magnitude diagram (CMD), in $\log (Z)$, and in height 
above the galactic plane.
Several comparisons with samples of stars classified by other means indicate
that for 4500 K $\leq \teff \leq 6500$ K, 
our classifications are reliable within about $\pm 200$ K and 
0.4 dex in $\log (g)$ for dwarfs, with somewhat larger $\log(g)$ uncertainties
for giants.
It is difficult to assess the reliability of our $\log(Z)$ estimates,
but there is reason to suspect that it is poor, particularly at
extreme $\teff$.
Comparisons between the CK models and observed colors are generally
satisfactory with some exceptions, notably for stars cooler than 4500 K. 
Of great importance for the {\it Kepler Mission}, 
for $\teff \leq  5400$ K, comparison with asteroseismic results shows
that the distinction between main-sequence
stars and giants is reliable with about 98\% confidence.  
Larger errors in $\log(g)$ occur for warmer stars, for which our filter
set provides inadequate gravity diagnostics. 
The KIC is available through the MAST data archive.
 
\end{abstract}

\keywords{catalogs --- methods: data analysis --- surveys --- techniques: photometric }

\section{Introduction}

The {\it Kepler Mission} \citep{bor10a} surveys some 1.6$\times 10^5$ stars in a 
field covering roughly 150 square degrees,
watching for short-lived dips in brightness that may signal transiting planets.
Of special interest to {\it Kepler} are transits by Earth-size planets,
which, if they involve Sun-size stars, give relative 
depths of
about $10^{-4}$, near to the practical limit of precision accessible by {\it Kepler}.
For a planet of given size, the transit depth scales inversely as the
cross-sectional area of the parent star.
For this reason, the detectability of Earth-size planets depends strongly on
the typical stellar radius of the sample of stars that {\it Kepler} observes.
The number of stars that {\it Kepler} can follow is limited by telemetry bandwidth,
and is considerably smaller than the total number of stars in {\it Kepler}'s field
of view (FOV) that allow useful photometric precision.
Thus, from an early stage the {\it Kepler} team recognized the importance of
characterizing the radii of stars in the {\it Kepler} FOV, 
to prevent large-radius stars (e.g. giants) from taking slots on the target
list away from smaller stars with better planet-detection prospects.
The project instituted the Stellar Classification Project (SCP) in response
to this need.
The SCP's goal was to provide, for all plausible target stars in the {\it Kepler}
FOV, estimates of important stellar parameters.  These were principally the
radius $R$, effective temperature $\teff$, and apparent magnitude
$K_p$ (ie, as seen by the $Kepler$ photometer) but also, to the extent possible,
the surface gravity parameter $\log (g)$  and the metallicity parameter
$[Z] \equiv \log(Z/Z_{\sun})$.

Because of the sky area and large number of stars involved, we deemed
spectroscopic classification to be impractical, and instead chose to use broadband
photometry, augmented by intermediate-band photometry using our 
custom $D51$ filter,
which is sensitive to surface gravity and to metallicity.
The results of this photometric reconnaissance of the {\it Kepler} FOV 
were federated
with other suitable catalogs, such as 2MASS \citep{skrut06}, 
USNO-B1.0 \citep{mon03}, Hipparcos \citep{sp1200, per97},
Tycho2 \citep{hog00}, and UCAC2 \citep{zac04} to become the {\it Kepler} Input Catalog, or KIC.
The aim of this paper is to describe how we carried out the photometric
analysis and stellar classification for the KIC.
Details of the observing routine and of the photometric reductions will be
given elsewhere, but a brief summary follows.

We took all observations with the 1.2m reflector at the 
Fred Lawrence Whipple Observatory
on Mt. Hopkins, AZ.
During the course of the project, we used a succession of three CCD cameras:
the {\it 4-Shooter} (from the project's first data in September 2003 = JD 2452895
until August 2004 = JD 2453233),
the $MiniCam$ until September 2005 = JD 2453626, and the $KeplerCam$ thereafter. 
All of these cameras had thinned, back-illuminated 4K x 4K pixel formats 
covering fields roughly
22 arcmin square, but the details of their detectors, noise properties,
sensitivity, and geometry varied from camera to camera.
To cover the entire {\it Kepler} field with these cameras required 1600 pointings
with minimal (roughly 10\%) overlap between adjacent pointings.
We began observations with 7 filters (nominal Sloan $u,g,r,i,z$, and two
special-order intermediate-bandwidth filters we termed $D51$ and $G_{red}$).
Both of the latter had bandpasses of about 15 nm.
The $D51$ filter, which was modeled after the Dunlap Observatory $DD51$
filter, was centered at 510 nm, and the $G_{red}$ filter at 432 nm.
In practice, we soon learned that the $u$ and $G_{red}$ filters 
required excessive
exposure times, so we took very few observations with these filters,
and we will henceforth ignore them for the most part.

For the great majority of the observations, we cycled through all of the
filters at one pointing before moving to the next pointing.
Each filter sequence consisted of both long 
and short integrations in the filters $g,r,i,z$.
For $g$ and $z$, the long and short exposure times were 30s and 3s; 
for $r$ and $i$ they were 20s and 2s.
Filter $D51$ got only a single long integration of 160s.
We identified 2 pointings containing stars that we used as secondary
photometric standards;  we returned to one of these fields about once per
hour, so as to have a fairly dense time sampling of the atmospheric
extinction.
We selected these two fields so that they included the open clusters
NGC 6811 and 6819; the centers of these fields lay at \{RA, Dec\} 
of
\{294.56, +46.58\} and \{295.28, +40.23\} (decimal degrees), 
and they contained 455 and
1181 secondary standard stars, respectively. 
Our intention, which we almost realized, was to visit each pointing at least 3
times under conditions of good transparency, never returning twice to a
given pointing (except the standards) on the same night.

We used a special-purpose pipeline to reduce the image data to catalogs
of star positions and apparent magnitudes (uncorrected for atmospheric
extinction).  This pipeline made the usual corrections for bias and
flat field, identified star-like objects, used DAOPHOT \citep{ste87} 
to perform PSF-fitting photometry
and computed aperture corrections based on isolated bright stars, and
fit an astrometric model to the stellar positions.
We fitted the stellar PSFs using the DAOPHOT ``Penny1'' function, which has
a Gaussian core and Lorentzian wings.
The pipeline returned a list of detected stars 
(and starlike objects such as radiation
events), with positions, magnitudes, sky background estimates, and shape
parameters for each.
Also returned were error estimates and various parameters relating to the
image as a whole, including the starting time, exposure time, and estimated
seeing width.
All of these data were saved in an ASCII file and passed to the software
that is the subject of the current paper.
Here we describe the methods used for photometric calibration,
correction for atmospheric extinction, and for
interpretation of the resulting absolute photometry in terms of the
physical parameters of the prospective {\it Kepler} target stars.

\section{Strategy}

The processes and software described here represent an intermediate
stage in the processing of SCP data for the KIC.
The functions of the procedure described here were fourfold.
First, it ingested the raw photometric catalogs provided by the photometry
pipeline into a group of databases
that allowed convenient processing.
Second, it estimated the atmospheric extinction suffered by each
measurement, and corrected the instrumental stellar magnitudes to yield
calibrated ones.
Third, it combined the calibrated magnitudes with other information
(e.g. stellar atmosphere models) to estimate the physical
parameters of each observed star, including $\teff$, $\log (g)$,
$[Z]$, radius $R$, mass $M$, and interstellar reddening $E_{B-V}$.
Last, it discarded those stars (and putative stars) for which 
estimates were deemed unreliable.

Data ingestion was a straightforward process, with its details determined
almost entirely by the database structure that was defined at the outset.
Correction for atmospheric extinction was also simple in concept, depending
mostly on the model adopted for extinction, and on the criteria for
estimating the parameters in that model.
Many of the latter decisions were guided by the choices made for the SDSS
\citep{fuk96, smith02}, since our filter set was similar to theirs.

Stellar parameter estimation was the most difficult part
of the project.
The underlying problem is that, for the purposes of the {\it Kepler} project,
the most interesting parameter is the stellar radius $R$,  
which is to say $\log (g)$. But the 
(mostly) broadband filters we used provide measurements that are almost 
entirely insensitive to this parameter.
The intermediate-band $D51$ filter provides some gravity sensitivity,
while for M-type stars, the $J-K$ color (obtained from the 2MASS catalog,
\citet{skrut06}) provides a gravity measure.
Nevertheless, the photometric information that we could obtain on the
timescale necessary for the project was barely sufficient for our needs.
It therefore made sense to perform the parameter estimation in the
context of astrophysical information external to our photometry.
We did this by adopting various distributions known for stars
in the Sun's neighborhood as Bayesian priors, and taking as each
star's physical parameters the ones that maximized the posterior
probability of obtaining the observed magnitudes and colors.
As amplified below, this method had several drawbacks, but it did
use the available data to far greater advantage than do methods that
ignore the prior constraints.
Many of the methods described below are devoted to ways of quantifying
the needed prior distributions.

Finally, filtering out bad data and formatting the result for the
next stage in KIC assembly was also straightforward,
consisting mostly of identifying spurious detections, failed fits,
and so on.
The criteria for passing results to the final KIC are described below.

\section{Calibrated Photometry}

The data received from the photometric pipeline consisted of one ASCII file for each
image taken at the telescope.
Each file contained a header and two data tables.
The header contained information relating to the image as a whole:
Julian date, telescope coordinates, filter, exposure time, as well
as various quantities such as PSF width, derived in the course
of the analysis.
The first data table contained
the stars' coordinates obtained from an astrometric
fit, their instrumental magnitudes, the local sky brightness, and various
error estimates and goodness-of-fit parameters.
The second table contained the coordinates, J, H, and K magnitudes, and
unique integer identifier for all of the stars in the 2MASS catalog 
\citep{skrut06} that
fell within the field covered by the image in question.

To allow simple comparisons with results from the SDSS, we wished to place
our photometry as nearly as possible on the Sloan system.
Unfortunately the {\it Kepler} field lies entirely at low galactic latitude,
so at the time we started the project, there were no SDSS data available
inside the {\it Kepler} field.
(This is no longer true, though SDSS coverage of the $Kepler$ field remains
very incomplete, with only a few square degrees of overlap in DR7.)
We therefore chose a set of 8 fields elsewhere in the sky,
each with data available from SDSS DR1 \citep{aba03, stou04},
to serve as our photometric standards.
Subsequent SDSS data releases, notably DR2, had improved photometry.
These improvements applied mostly to galaxies, however.
Changes in methods for fitting ``model'' magnitudes gave modest reductions in
the scatter for stellar magnitudes, but left the mean zero points unchanged
\citep{aba04}.

We chose the standard fields first to span a range of RA surrounding 
the {\it Kepler}
field, and also so that each contained a fairly wide range of stellar colors
(but even so, there were many fewer blue than red stars in the standard
fields).
There were 284 primary standard stars;
ignoring 6 evident outliers (which varied from the mean $(g-r),(r-i)$ 
color-color relation by more than 0.5 magnitude), 
they spanned the range $-0.4 \le (g-r) \le 1.4$,
and $-0.2 \le (i-z) \le 0.4$ (although the great majority of them lay
within $0.3 \le (g-r) \le 0.8$).
We observed several standard fields each night for several months
at the beginning of the
project, and used their time-averaged, extinction-corrected magnitudes
to fit transformation coefficients between our internal magnitudes and
the SDSS system.
Subsequent visits to the standard fields provided assurance
that our photometry was consistent.

Because of their source, primary standards did not have observed magnitudes
in the D51 filter.
To define the instrumental magnitude scale in this filter, we set the
D51 magnitude for these stars so as to force agreement with
model $(g-D51)$ colors, given the stars' observed (SDSS) 
$(g-r)$ colors.
We computed the model magnitudes used for this purpose from the Basel 
models \citep{lej97} with $[Z]=0$ and $\log(g) = 4.0$, and from the estimated
wavelength response for the 1.2m telescope 4-Shooter camera.
(We did not use the CK models for this purpose because they were not
available at the early stage of the project when we did this calibration.)

The supplementary material contains a table listing all 284 of the primary
standard stars, along with their photometric indices and the stellar
parameters that we inferred for these stars using the methods described
below.
The first few lines of this compilation are displayed in Table 1.

Later in the project, it was desirable to build reference fields
of secondary standard stars.
The stars in these fields had their magnitudes defined by repeated calibration
against the primary standards.
Two of the pointings that we used for secondary standards were chosen 
within the {\it Kepler} field itself.
In the normal observing sequence, the telescope returned to one of these
pointings on an hourly basis, so that we could obtain reliable measures of
extinction in the part of the sky that was of most interest.
Other secondary standard fields were defined near objects 
(for example, the cluster M67) that we wished
to use in the astrophysical calibration of the photometry.

After obtaining a few dozen nights of data with all of the secondary 
standard stars,
some stars could be identified as unsuitable because they showed large
temporal variability, or they were members of close binaries, or for similar
reasons.
We removed these from the list of secondary standard stars.

\subsection{Estimating Atmospheric Extinction Parameters}

The extinction model was based on that used for the SDSS standard star
grid \citep{smith02}, although in practice we found several of the coefficients
in this model to be difficult or impossible to measure on a
nightly basis, and we therefore set them to constant (sometimes zero) values.
We partitioned our photometric observations into time-contiguous 
units termed $blocks$,
with a block being a unit of data that could be adequately described
by a single set of extinction coefficients.
Almost always, blocks corresponded to entire observing nights.
Sometimes, however, changing weather conditions made it desirable to subdivide
a night into multiple blocks.

For filter $i$ and an instance $j$ (within a block) 
of observing a particular star $k$, 
our extinction model represented the observed magnitude 
$m_{ijk}$ 
as follows:
$$m_{ijk} \ = \  m_{0ik} \ + \ a_i \ + \ k_i(X_{kj} - X_0) \ + \ $$
$$b_i(C_{ik}-C_{0i}) \ + \  c_i
(C_{ik}-C_{0i})(X_{kj} - X_0) \ \ . \eqno(1)$$
Here $m_{0ik}$ is the true magnitude of star $k$ in filter $i$,
$X_{kj}-X_0$ is the difference between the airmass of star $k$ at
instance $j$ and a standard airmass $X_0$,
$C_{ik}$ is the color of star $k$ defined using a particular pair of filters
similar in wavelength to the filter $i$, 
and $C_{0i}$ is the color of a 'typical'
star using that same filter pair.
The coefficients $a_i$, $k_i$, $b_i$, and $c_i$ are parameters that
may be chosen to give the best fit to the observations.

Note that by writing the extinction as above, the 'standard' magnitude
of a star corresponds to what one measures when the star lies at airmass $X_0$,
not when it is outside the atmosphere.
One advantage of this approach was to reduce the effect of errors in
estimation of the extinction per unit airmass (the $k_i$).
Also, this choice of airmass coordinate tends to reduce correlated errors
between $a_i$ and $k_i$.
For this analysis, we took the standard airmass $X_0$ to be 1.215.
We chose this value to coincide with the value 1.3
adopted by SDSS, with allowance for the higher altitude (hence
lower air density) of the Sloan telescope relative to the 1.2m
telescope at Mt. Hopkins.

Note further that the extinction model makes no explicit assumption about,
or measurement of, the site's mean extinction as a function of wavelength.
All information about this behavior is contained in the coefficients
$a_i$ and $k_i$.
As will be shown below, these vary enough from night to night that they call
into question the utility of the site's mean color-dependence of extinction.

In practice, we found that the observations of standard stars on a single
night ordinarily did not suffice to give a reliable estimate of all of
the coefficients in the extinction relation.
Indeed, we obtained best results by fitting for the coefficients $a_i$ each
night, using values averaged over an entire observing season for the 
coefficients $k_i$, and taking the coefficients $b_i$ from theoretical
calculations based on model stellar fluxes and the estimated wavelength
responses of the various filters.
We took the coefficients $c_i$ to be uniformly zero.
Figure 1 shows the time variation of the coefficients $a_i$ for all but 2 of the
205 nights for which we have data.
The zero points on this Figure have been arbitrarily shifted for plotting
clarity.
The zero points are arbitrary, but time variations of the $a_i$ coefficients
have physical interest.
Significant jumps in the $a_i$ coefficients occurred with the inauguration of new
cameras (changing from 4-shooter to MiniCam about JD 2453233, and from MiniCam
to KeplerCam about JD 2453626).
There also are trends and seasonal variations visible in these
data, notably a loss of about 20\% sensitivity in the telescope/KeplerCam 
system since the time the KeplerCam was installed.

After estimating the $a_i$ coefficients, we produced a second
estimate of the extinction and the quality of the night's data.
For this purpose we used only selected stars observed at each reference pointing
to estimate the $a_i$ and $k_i$ coefficients for each night.
We chose stars so that exactly the same subset of stars in each reference
pointing were always used in the fitting process for a single night.
This minimized errors arising from errors in the assumed magnitudes of the
reference stars.
We thus obtained nightly estimates of the $k_i$ coefficients
as well as the $a_i$ (although the $k_i$ are well-determined on a little less
than half of the nights).
Figure 2 illustrates such a fit for one fairly typical night.
Frequent returns to reference fields within the $Kepler$ FOV allow accurate
fits for $k$ on photometric nights, and reveal the times and severity of
time-varying extinction on the (more common) non-photometric nights.
The bottom panel shows residuals around the fits plotted against the
measured image Full Width at Half Maximum (FWHM), measured in arcsec.
An important part of the photometric reduction process was to perform
seeing-dependent corrections for the fraction of starlight falling outside
the boundaries of the effective photometric aperture.
On most nights, this plot showed no clear trend of residuals vs FWHM,
but on some (such as the one illustrated here), there is slight 
evidence for such a trend.
This indicates occasional errors in the aperture-correction procedures,
which must compromise the photometry at some level.
This problem was sporadic and relatively minor, making it difficult to
diagnose.
Indeed, we never succeeded in tracking this problem to its root, though
plots such as those in Figure 2 allowed us to identify problematic
nights.
Also evident from this plot is the relatively large FWHM of the images
on the night displayed here.
Indeed, the images were broad on a large fraction of our
observing nights;  the image FWHM was below 1.7 arcsec only about 2\%
of the time, and the median FWHM for all of our images was 2.5 arcsec.
Because of this relatively poor spatial resolution, the KIC is ineffective
at distinguishing the components of binary stars, if their component
separation is even moderately small.

Figure 3 shows the variation of the $k_i$ with time, measured on the nights
when we judged the observations to be consistent enough to allow a measurement.
The night-to-night variations tend to be large (up to 40\% of the mean values),
but they are also well-correlated among filters.
Thus, there is good evidence for variations dominated by large-particle
aerosol extinction, having
little wavelength dependence, and varying on a time scale of one to a few days.
The SCP photometry could be improved by putting this information about the
$k_i$ back into the extinction model on a nightly basis.
We have not yet done this, however.
The corrections would not be large, in any case.
Given the size of the errors in $k$ and the typical difference between the
airmasses of actual observations and the standard airmass ($X_0 = 1.215$),
corrections in the observed magnitudes would be at most 0.09 mag, and would
be less than .03 mag for about 90\% of the observations.

After estimating the extinction coefficients for a block of data, we 
estimated the $m_{i0}$ magnitudes averaged over visits
 for every star that was observed
in any filter within that block.
This process identified every star that was observed at any time during
a given block, and then gathered together every observation to be found in the
databases for each of those stars.

Of the 1600 pointings that span the entire {\it Kepler} field, none have zero visits
with acceptable photometry,
8\% were visited only once, 71\% twice, 17\% three times, and 3\% four or more
times.
Two pointings, corresponding to secondary standard fields, were visited more
than 260 times each.
For pointings with at least 4 visits,
we computed a pseudo-$RMS$ from the interquartile dispersion
$q$ as $RMS = q/1.349$;  this formula gives the expected result for a gaussian
distribution, but is insensitive to a small proportion of extreme outliers.
we identified outliers as measurements differing from the median by more than
5 times this pseudo-$RMS$, and discarded them from the fit to give robust 
minimum-$\chi^2$ magnitude
estimates.

\subsection{Precision of Corrected Magnitudes}

We assessed
the precision (in the sense of repeatability) of stellar magnitude estimates 
by collecting all the measurements for each star in the secondary
standard fields (each having typically hundreds of individual observations),
and computing the scatter in these time series.
We did this for a set of 5652 stars brighter than $r = 19.5$, each
having at least 50 individual
observations (and most having more than 200).
For each star we estimated the time-series pseudo-$RMS$ as described above.
We then computed for each filter the median pseudo-$RMS$ taken over stars, in 
1-magnitude bins centered on integral
values of the (time-series) median magnitude.
The resulting pseudo-$RMS$ scatter as a function of stellar magnitude 
in each filter is plotted in
Figure 4. 
The photometric errors are dominated by brightness-independent
processes (probably errors in the atmospheric extinction correction)
in all filters for stars brighter than about magnitude 14;  for fainter
magnitudes, photon statistics begin to have an effect, and become dominant
by about magnitude 16.
For the stars brighter than magnitude 14, the repeatability of 
extinction-corrected measurements
is about 2\%, almost independent of filter.
 
The RMS precision of color estimates 
(ie, differences between magnitudes measured
in different filters) was usually somewhat better than for magnitudes (typically
1.5\%), because much of the scatter in the extinction-corrected magnitudes
comes from extinction processes that have long time scales, whereas the
two measurements making up a color estimate were almost always taken within
a few minutes of each other.

\section{{\it Kepler} Magnitudes}

To estimate planet detectability for each potential target star, the {\it Kepler Mission} required information
about stellar magnitudes as they would be measured by the $Kepler$ photometer.
These are known as the $Kepler$ Magnitudes, or $K_p$.
For this purpose we estimated $Kepler$ Magnitudes  using photometry in standard
bandpasses. 
The $K_p$ values that we computed for each star are tabulated
in the KIC along with the other photometry.
The ideal $Kepler$ wavelength response function $K_I(\lambda)$ may be 
found in \citet{koc2010}
and on the $Kepler$ web site \footnote{http://keplergo.arc.nasa.gov/CalibrationResponse.shtml}.
The bandpass may be roughly described as having sharp edges, with the center
being slightly peaked; it has more than 10\% response in the wavelength range
$420 \ \leq \lambda \leq \ 890$ nm.
Thus, the bandwidth is about 470 nm and the effective wavelength, for an SED
corresponding to a 5500 K black body, is approximately 665 nm.

The $K_p$ are defined as $AB$ magnitudes \citep{oke74, smith02}, 
derived from each target's
calibrated ${g,r,i}$ magnitudes.
To compute them, we started with the published \citep{cas04} grid of 
stellar atmosphere model fluxes, and used these and the 
known, tabulated wavelength response
functions to compute the rate of photoelectron detections from each of the
model flux distributions for a wide range of stellar types
(these were functions of $\teff$, $\log(g)$, and 
metallicity) using each of
the filters $\{g,r,i\}$ and using the ``ideal'' $Kepler$ 
bandpass $K_I$ described above.
We then attempted to approximate these synthetic magnitudes as 
they would be measured
in the band $K_I$ in terms of linear 
combinations of the $\{g,r,i\}$ magnitudes for the corresponding models.
Based on a visual inspection of the relationships, we defined these combinations
on each of several contiguous ranges of some fiducial color, e.g. $(g-r)$;
for operational purposes, $K_p$ is defined only in terms of these 
combinations of magnitudes in standard filters.

Complications arise because not all KIC stars have valid values for all
of the filters $\{g,r,i\}$, and moreover many stars (which appear in
the KIC by virtue of federation with non-SCP catalogs) have none of them.
The $Kepler$ Magnitude $K_p$ is thus defined by a complex set of rules,
depending on what information is available about the star in question.
Policy demanded that the KIC contain a Kepler magnitude for every star known to
lie within the field.
Thus, the rules covered cases in which information was scant and accurate transformations were not possible.

We note that the definition of $K_p$ as a linear combination of observed
magnitudes in standard bands means that the effective wavelength response 
function that one should ascribe to $K_p$ is ill-defined.
Apart from the ambiguity arising from which bands may have been used in
computing $K_p$, it develops that (because of the nonlinear relation between
flux and magnitude) the effective bandpass for $K_p$ is also a function of the
stellar color.
Thus, the magnitudes $K_p$ are entirely defined by the following Equations 
(2 through 6).
A further result of the method for computing $K_p$ is that the photometric zero
point for $K_p$ is not exactly on the AB system but is rather a weak function 
of stellar color.
Because we chose the weighting coefficients for the various bandpasses 
based on a color-dependent fit to integrals over $K_I(\lambda)$, the part
of the zero point dependence that is linear in color has been absorbed
by the weighting coefficients.
Higher-order terms remain, however.
At worst, for stellar colors at the extreme ends of the ranges that one
encounters, these higher-order terms amount to about 0.05 mag.

For stars with SCP photometry, the computation rules are as follows.

If only one of $\{g,r,i\}$ is valid, then
$K_p$ is equal to the valid magnitude.

If only $g$ and $r$ are valid, then
$$K_p \ = \ 0.1 g \ + \ 0.9 r, \ \ \ (g-r) \leq 0.8, \eqno(2a)$$
$$K_p \ = \ 0.2 g \ + \ 0.8 r, \ \ \ (g-r) > 0.8. \eqno(2b)$$

If only $g$ and $i$ are valid, then
$$K_p \ = \ 0.55 g \ + \ 0.45i, \ \ \ (g-i) \leq -0.05, \eqno(3a)$$
$$K_p \ = \ 0.3 g \ + \ 0.7 i, \ \ \ (g-i) > -0.05. \eqno(3b)$$

If only $r$ and $i$ are valid, then
$$K_p \ = \ 0.65 r \ + \ 0.35 i, \ \ \ (r-i) \leq 0.673, \eqno(4a)$$
$$K_p \ = \ 1.2 r \ - \ 0.2 i, \ \ \ (r-i) > 0.673. \eqno(4b)$$

If $\{g,r,i\}$ are all valid, then
$$K_p \ = \ 0.25 g \ + \ 0.75 r, \ \ \ (g-r) \leq 0.3, \eqno(5a)$$
$$K_p \ = \ 0.3 g \ + \ 0.7 i, \ \ \ (g-r) > 0.3. \eqno(5b)$$
 
The linear combinations in Eqns. (3), (4), and (5) 
yield $K_p$ magnitudes that reproduce the ``ideal''
magnitudes $K_I$ with typical errors of about 0.03 mag for stars with
$\teff \ge 3500$K, though for extreme stellar parameter values, the errors may
reach $\pm$ 0.2 mag.
In the case in which only $g$ and $r$ are known (Eqns. 2), the
disagreement between $K_p$ and $K_I$ may reach 0.6 mag for very cool M-type
stars, in the sense that the computed $K_p$ is fainter than $K_I$.
  
For stars imported from the Tycho-2 parent catalog, 
which gives $B_T$ and $V_T$
magnitudes, we estimate equivalent Sloan magnitudes $g_T$ and $r_T$ as
$$g_T \ = \ 0.54 B_T \ + 0.46 V_T- 0.07, \eqno(6a)$$
$$r_T \ = \ -0.44 B_T \ + \ 1.44 V_T +0.12, \eqno(6b)$$
and then compute $K_p$ using Eqns. (2a), (2b).

For stars coming from any other parent catalog having a both blue and a red
magnitude,
we simply substituted these magnitudes for $g$ and $r$ respectively,
and again used Eqns. (2a), (2b).

Finally, for stars coming from parent catalogs that contain information
in only one optical bandpass, $K_p$ is taken to be the reported magnitude
in that bandpass.
In this case, of course, very large errors (1 mag or more) may occur.
 
\section{Model Stellar Atmospheres}

The stellar classification process requires choosing among model stellar atmospheres
(which have parameters $\{ \teff, \log (g), \log (Z) \}$) so as to best fit the
photometric observations.
We used model atmospheres by \citet{cas04};  
these cover $3500K \leq \teff \leq 50000K$,
$0 \leq \log (g) \leq 5.5$, and $-3.5 \leq \log (Z) \leq 0.5$, although not all
gravities are represented at all temperatures.

The models provide fluxes as a function of wavelength.
To convert these to stellar AB magnitudes (relative to an arbitrary zero point)
we transformed the tabulated values from units of energy flux per unit
wavelength to units of
photon flux per unit frequency, multiplied the stellar flux by an estimate of 
the CCD response and by
the estimated filter transmission functions, 
and integrated the result over frequency.
to give the rate of photoelectron production as a function of filter and of each
of the parameters that characterize the model atmospheres.
We took magnitudes in each of the filters to be on the AB magnitude scale.
In operational terms
this was already the case for the filters $\{g,r,i,z\}$ and for the 2MASS
magnitudes $\{J,H,K\}$; we therefore left the zero points in all of these
filters untouched.  For the D51 filter, we adjusted the zero point to
force agreement with the
CK model $g-D51$ colors as described in section 3. 

After computing the stellar magnitudes described above,
we next formed differences to yield 7 independent colors, viz.,
$(g-r),(r-i),(i-z),(z-J),(J-H),(H-K),(g-D51)$.
The arbitrary zero point cancels from these differences, so these numbers
are a representation of the spectral energy distribution
of the stellar models that does not depend upon stellar radii or distances.
Most of the operations that later tried to match stellar properties to observations
relied on these 7 colors.

To achieve near equality between observed colors and fluxes from the CK
models, we transformed the latter in several ways.
First, we adjusted the magnitude zero points for all filters except $i$ (which
we adopted as standard, linked by definition to a small set of SDSS stars).
We based these adjustments on comparisons with 3 sets of stars for which
the photometric (and in some cases the physical) properties could be known 
with some accuracy from pre-existing data.
These included a set of 4 Sun-like stars chosen from the list of SDSS 
standard stars \citep{smith02},
a set of 63 to 77 stars (depending on the filter involved) in the cluster
M67, each of them known to be single from spectroscopic observations
by W. Latham and S. Meibom, and having good photometry and
spectroscopic classification \citep{mon93,san04},
and finally a set of some 500 probable cluster stars in a 2-degree
square surrounding M67.
For the last group, we first identified an optimally-fitting 
stellar atmosphere model for
each star that fell close to the M67 cluster isochrone,
and then adjusted the magnitude zero points to minimize
the mean difference between observed and predicted colors.

Second, we noted a serious disagreement between observed and model $(g-r)$
colors for stars cooler than 4000 K.
Model $(g-r)$ colors based on the CK fluxes never exceed about 1.2,
while many faint members of M67 and nearby field stars
(which are almost certainly dwarfs of near-solar metallicity) have $(g-r)$
colors that approach 1.5.
To allow more accurate modeling of the photometry for these stars,
we applied ad hoc adjustments to the $(g-r)$ colors of all CK models
with temperatures of 4000K or less.
These adjustments added
to each model $(g-r)$ value an offset that depended only upon $\teff$.
The corrections applied were +0.179 at 4000K, +0.289 at 3750K, and +0.402
at 3500 K.
This ad hoc process improved the quality of fits considerably,
but rendered the $\teff$ scale in this range 
unphysical, at least so far as the $(g-r)$ color is concerned.

Third, we added smaller $\teff$-dependent corrections to the colors 
$\{ (r-i),(i-z),(J-H),(H-K) \}$.
These corrections were linear in temperature and 
equal to zero at $\teff = 5000$K,
with slopes per 1000K equal to $\{.006,-.021,-.036,.011 \}$ respectively,
.
All of these adjustments were independent of $\log (g)$ and $\log (Z)$.

Last, we adjusted the (g-D51) color in a way that depended both
on $\teff$ and on $\log (g)$:  for $\teff \leq 4300$K,
the revised model color $(g-D51)^{\prime}$ was given by
$$(g-D51)^\prime \ = \ (g-D51) \ - \ 0.23 f_g (\teff-4300)/1000. \eqno(7)$$
where the function $f_g (\log (g))$ was unity for $\log (g) \geq 3.5$, 
and for smaller $\log(g)$ decreased linearly
with $\log (g)$ to zero at $\log (g) = 0$.

\section{Reddening}

Interstellar reddening is significant for most of the stars in the {\it Kepler}
field, so it was necessary to include it in the models of stellar colors.
We assumed that the wavelength dependence of reddening is described
by the formalism of \citet{card89}.
In all cases we used $R_V \ \def \ A_V/E_{B-V} = 3.1$, which Cardelli et al.
found to be the typical value for diffuse interstellar dust clouds.

Strictly, the reddening suffered by starlight depends on the spectrum
of the star, and hence on its parameters $\{\teff, \log (g), \log(Z)\}$.
The importance of this effect is fairly small, however, so for speed
of computation, we precomputed reddenings for our filters and
a range of stellar parameters.
We then adopted the reddening vectors for typical stars in the {\it Kepler}
field ($\teff=5000$K, $\log (g) = 4.0$, $\log (Z) = 0$) as applying
to all stars.

We estimated the strength of the reddening using
a simple model of the dust distribution in the Milky Way.
This model assumed that dust is distributed in a smooth disk
aligned with the plane of the Milky Way, having an exponential
decay of density with height above the plane.
For all computations described here (except for certain test cases,
such as parameter estimations for M67), we took the e-folding height
for disk density to be 150 pc, which is slightly larger than suggested
by recent estimates (\citet{kop98, mar06} give 140 pc, 125 pc, respectively).
We took the density in the plane to be
such as to cause 1 magnitude of extinction in the $V$ band in a path
length of 1000 pc \citep{kop98}.

\section{Fitting Stars to Observations:  Linked Parameters}

A star and the light we see from it may be characterized by many parameters,
not all of them independent.
These include not only its photometric colors, but also its apparent
magnitude, distance, reddening, and intrinsic properties $\{ \teff,
\log (g), \log(Z) \}$, mass $M$, radius $R$, and luminosity $L$. 
Different photometric measurements constrain different combinations of
these parameters;
to obtain sensible estimates of the intrinsic parameters, it is important
to respect the relations that connect them.

We took our distinct observables to be the photometric colors
(in the typical case for which $u$ and $G_{red}$ are unavailable, 
there are 7 of these), the apparent magnitude
in any one filter (we used $r$), and the galactic latitude $b$.
The eleven parameters we wished to estimate were $\teff, \log (g), \log (Z),
M, R, L, BC_r, d, A_r, A_V, E_{B-V}$, where $BC_r$ is the bolometric correction
for the $r$ band (in practice computed as $BC_r = BC_V + (V-r)$), $d$ is the distance in pc, $A_r$ and $A_V$ are the 
interstellar extinction in the $r$ and $V$ bands, and $E_{B-V}$ 
is the $B-V$ color excess due to interstellar extinction.
But connecting these parameters are the relations:
$$A_r \ = \ \kappa_r \int_0^d \exp (-s \sin b) ds \eqno(8)$$
$$A_V \ = \ A_r / 0.88 \eqno(9)$$
$$E_{B-V} \ = \ A_V/3.1 \eqno(10)$$
$$L \ = \ R^2 (\teff / T_{\sun} )^4 \eqno(11)$$ 
$$r \ = \ r_{\sun} \ - \ 2.5 \log L + \ BC_r \ + \ A_r \ + \ 5 \log d  - 5 \eqno(12)$$
$$\log (g) \ = \ \log (g)_{\sun} \ + \log M \ - \ 2 \log R  \ \ ,\eqno(13)$$
where $\kappa_r$ is the assumed $r$-band interstellar extinction coefficient in
magnitudes per pc, 
$R$, $M$, and $L$ are in solar units, $d$ and the dummy integration variable
$s$ are in pc, 
$r_{\sun}$ is the Sun's absolute magnitude in the $r$ band,
and $A_r$, $A_V$,
$BC_r$, and $E_{B-V}$ are in magnitudes.
The simple numerical relations between $A_r$, $A_V$, and $E_{B-V}$ result from
integrating the Cardelli et al. model of interstellar reddening against
typical (5000K dwarf) stellar flux distributions, as described above.

Applying the above constraints reduced the number of unknowns from 11 to 5, but
this was still an uncomfortably high-dimensional space to search for best
goodness-of-fit.
Worse, given realistic errors in the data, we found that 
an unrestricted search in this space
tended to lead to unphysical solutions in which various parameters were driven to
extreme values in an attempt to balance misfits relative to the observations.
To reduce these problems we adopted two further ad hoc relations, namely
$$BC_r \ = \ BC_r(\teff, \log (g)) \eqno(14)$$
and
$$M \ = \ M(\teff, \log (g)) \ \ .\eqno(15)$$
The $BC_r$ relation is valid for fixed composition, and for computational purposes
we simply adopted the results of the CK models for $\log (Z) =0$.
We justify the inaccuracies resulting from ignoring the $\log (Z)$ dependence 
by the relative scarcity of low-metallicity stars in the solar neighborhood.

The $M(\teff, \log (g))$ relationship is true only in a statistical sense,
because age and (to a lesser extent) metallicity effects cause the evolutionary
tracks of stars of different mass to cross in the CMD.
This causes inevitable ambiguity, especially for giants and subgiants.
However, for the cool main-sequence stars that are of the most interest to
{\it Kepler}, the approximation is fairly good.
To estimate this relation, we took stellar evolution isochrones by \citet{gir00},
and weighted each by its age (thus, in effect, assuming a constant star formation
rate), 
with 5 isochrones covering ages between 63 My and 4.5 GY.
We then smoothed the resulting distribution of masses in $\log (\teff)- \log (g)$
space to yield typical masses at each point.

Having adopted the $BC_r(\teff, \log (g))$ and
$M(\teff, \log (g))$ relationships, we saw that
at the same level of consistency, several other useful parameters of stars
could also be treated as being functions only of $\teff$ and $\log (g)$.
These included the luminosity $L$, the $V-r$ color (used in relating extinction
to distance), and absolute magnitudes in $V$ and $r$ bands.
Thus, we precomputed tables for all of these quantities, and 
interpolated into the tables as necessary to give values for any of 
these quantities
as functions of $\teff$ and $\log (g)$.

The process of computing a goodness-of-fit statistic
for an individual star then proceeded as follows.
We first fetched the observed colors, the $r$ magnitude, and the galactic
latitude $b$ for each star.
We then performed a straightforward (ie, not particularly efficient)
search in $\{ \teff$, $\log (g)$, $\log (Z) \}$, 
seeking the minimum value of a merit
function that is the sum of a $\chi^2$ statistic based on the photometry and
the negative logarithm of a prior probability.
This section describes the $\chi^2$ computation;  the justification for this general
Bayesian procedure and the computation of the prior probability distribution will be
described below.
The search for the minimum involved evaluating the merit function on 
successive 2-dimensional
subsets of the search space;  for practical purposes this meant evaluating model
colors for a grid of models, each with specified $\teff$, $\log (g)$, and
$\log (Z)$, and each relating to a star with a specified $r$ magnitude and galactic
latitude.
We did this by using an interpolated table look-up 
to give $L$, $BC_r$, and $V-r$ from $\teff$ and $\log (g)$,
and then computing the distance $d$ and extinction from $r$, $b$, and the
galactic extinction model. 
We then generated model colors for
each $\teff$, $\log (g)$, $\log (Z)$ in the current grid of interest,
and, knowing the extinction for each of these, applied a corresponding
reddening correction to the model colors.
Last, we differenced the observed and model colors, and (using the estimated
observational errors) computed a $\chi^2$ value for each grid point.

For some stars, not all of the ${g,r,i,z,D51}$ data were available.
In these cases, we set the uncertainty for the missing data to $10^5$
magnitudes, assuring that the actual values made no contribution to the fit.
 
\section{Bayesian Posterior Probability Estimation}

When fitting stellar parameters to our photometric data, we found that
pure $\chi^2$ minimization 
often led to unreasonable results.
This is because outlandish combinations of the stellar parameters, corresponding
to stars that are seldom or never seen in nature, may yield marginally better
fits to the observations than do more plausible parameter combinations.

Bayesian methods provide a way to control this behavior.
To describe the application of Bayes' Theorem to the stellar 
classification problem, we denote
the intrinsic stellar parameters $\{ \teff, \log (g), \log (Z)\}$
by the vector ${\bf x}$, and the various photometric observations of a
particular star by the vector ${\bf q}$.
We assume that we know something about the statistical distribution of
stars, so that it is meaningful to talk about the prior probability distribution
$P_0({\bf x})$, where the probability that a given star lies within 
a small volume of
parameter space is given by $P_0({\bf x }) d{\bf x}$.
Then Bayes' Theorem says that if we add to this a priori information about stars
a set of observations ${\bf q}$ relating to a particular star,
then the updated (posterior) probabilities are
$$P({\bf x} | {\bf q}) \ = \ [ P_0({\bf x}) P({\bf q} | {\bf x})] / P_0({\bf q}) \ \ .\eqno(16)$$
The choice of stellar parameters that maximizes the posterior probability
is then the choice of ${\bf x}$
that maximizes this expression.
Here the denominator, the a priori probability of observing photometric
indices given by ${\bf q}$, is a normalization that does not depend upon the parameters
${\bf x}$ of interest;  for purposes of maximizing the posterior probability,
it may be ignored.
In the ideal case of independent Gaussian errors in the photometric observations,
one can write
$$P({\bf q} | {\bf x}) \ = \ \exp (-\chi^2 / 2) \ \ ,\eqno(17)$$
where $\chi^2$ (not the reduced $\chi^2$) is the usual goodness-of-fit statistic.
Taking the logarithm of the posterior probability, it follows that maximizing
this probability is equivalent to maximizing
$$ \ln P({\bf x} | {\bf q}) \ = \ \ln P_0({\bf x}) \ - \ \chi^2 /2\ \ .\eqno(18)$$
This is what the stellar parameter-estimation procedures try to do.

It is well known that Bayesian methods have both advantages and disadvantages.
In the current case one advantage is that wildly implausible stellar classifications
are ruled out a priori;  the emerging values of $\{\teff, \log (g), \log (Z) \}$
are guaranteed to resemble those of known kinds of stars.
The corresponding disadvantage is that stars with rare properties are almost
certain to be misclassified, since such stars are represented insignificantly
or not at all in the statistical samples that one uses to estimate
$P_0({\bf x})$.
The Bayesian maximization places stars in the most plausible of the already-
existing pigeonholes;
if the data are poor or contradict expectations for known kinds of objects,
incorrect classifications may occur.

Thus, for the principal {\it Kepler} purpose of distinguishing GKM-type 
dwarfs from giants,
the Bayesian approach can be expected to work well.
For other purposes (identifying possible brown dwarfs, say, or
distant highly-reddened blue supergiants), it is better to rely directly on the
photometry, and ignore the classifications.

Another problem with Bayesian estimation is that commonly there is no good basis
for estimating prior probabilities.
Fortunately, in the case of stellar classification, one can do rather well.
For purposes of the KIC classification, we expressed $P_0({\bf x})$ as the
product of 3 terms:
$$P_0({\bf x}) \ = \ P_{CMD}[\teff,\log (g)] P_Z[\log (Z)] P_z(z) \ \ ,\eqno(19)$$
where $P_{CMD}$ describes the probability density of stars in the
$\teff-\log(g)$ plane, and $P_z$ gives the probability density with height
above the galactic plane $z$.
Expressing $P_0$ as a product distribution implicitly assumes that there are
no correlations among the various dependencies.
Of course, this is not completely true.
Because of the relations between age and $Z$, and between age and $z$,
the product distribution is not strictly valid.
The relatively low frequency of old, low-metallicity stars makes this distribution a
reasonable approximation, however. 
Nevertheless, as we have implemented it, the
classification scheme knows nothing of (say) the existence of 
old low-metallicity halo
stars, nor of the absence of young low-metallicity stars, etc.

To estimate $P_{CMD}$, we used the Hipparcos (\citet{sp1200}; \citet{hog00}) 
catalog to create a histogram
of the nearby star distribution, sampled in $(B-V)_{Tycho}$, $V_{abs}$ space.
For this purpose we took all the Hipparcos stars with parallaxes that are known
to better than 10\%.
This sample contained 9590 stars.
We then mapped the $B-V$ color and absolute
$V$ magnitude into $\log (\teff)$ and $\log (g)$.
Once each star was associated with its calculated $\teff$ and $\log (g)$, 
it was a simple
matter to construct a histogram giving the fraction of stars 
found within each cell
in that space.

After creating the histogram of star frequencies in $\teff-\log (g)$ space,
we performed several edits and additions to make it more realistic,
complete, and useful.
We first set star frequencies to zero for regions where only 1 or 2 stars were found in
isolated cells.
We next added stars to the area of the histogram occupied by bright giants,
since the volume covered with good parallax precision by Hipparcos is not large
enough to populate this part of the diagram.
We then scaled the frequencies for intrinsically faint stars to account for the
search volume that is implied by the stellar absolute magnitudes,
since this volume is determined by the astrometric precision only for relatively
bright stars.
After doing this, the very faintest M stars were still unrepresented, so we added
stars to populate the faint tail of the main sequence.
Finally, we smoothed the histogram to yield one that represents the broad features
of the local stellar population, but that has little small-scale structure.
We took the natural logarithm of this histogram
to be the logarithm of $P_{CMD}$.

To estimate $P(Z)$, we used the compilation by \citet{nord04} of metallicities for
about 14000 bright nearby stars.
We first used these data to construct a histogram of $\log$(relative frequency) 
at equally-spaced intervals in $\log (Z)$.
We then fit a polynomial to these histogram values, and made a linear
extrapolation of the polynomial for $\log(Z) \leq -3$ and for
$\log(Z) \geq +0.9$.
We then evaluated this extrapolated polynomial to compute our estimate of 
$P(\log(Z))$.

For the distribution of stars as a function of the height $z$ above the
galactic plane, we assumed 
the galactic disk's number density to decrease exponentially
with a vertical scale height of 300 pc (\citet{cox00} p. 482).
In the probability distribution, we also allowed for the cubic increase 
in sample volume (per unit solid angle and apparent stellar magnitude) 
with increasing distance.

With these probabilities in hand,
we were able to compute the natural logarithm of the
prior likelihood $\ln (P_0 ( {\bf x}))$, according to Eqn. (19).

\section{Star Selection, Formatting, and Output}

When constructed as just described, the KIC contained a large fraction
of spurious objects.
These were mostly photometric artifacts that
appeared only once in all of our (multi-filter, multi-visit) catalogs.
The causes for these included near-threshold noise detections, radiation
events, diffraction artifacts near bright stars, ghost images caused by
crosstalk among the CCD quadrants, residual cosmetic defects in the CCDs,
and perhaps other sources that we have been unable to identify.
To remove such invalid data from the KIC, we discarded all supposed stars 
that did not have at least 2 valid optical measurements or 3 NIR measurements.
This rejection step removed a large fraction (about $2/3$) of the entries
from the star catalog.
Although this fraction appears very large, one must recall that each
pointing accumulated spurious detections (eg, radiation events) 
from images taken with every filter and at every visit, 
whereas the inventory of real stars
remained nearly constant.
Thus, the secondary standard pointings, which we visited hundreds of times
each, were prolific sources of false entries in the initial catalog
of detected objects.

Next, we set the astrophysical parameters $\{ \log (g), \log (Z),
A_V, E_{B-V} \}$ 
to be ``invalid'' for all stars where the number of valid measurements in 
visible-light filters
was less than 4, or where the stellar parameter estimation code
did not converge.
This step typically caused only a small fraction of the remaining stars
to be assigned ``invalid'' physical parameters.
We set the parameter $\teff$ to ``invalid'' only if there was no valid
optical color, which is to say, 2 or fewer valid visible-light magnitudes.

For a description of all of the data fields contained in the KIC, see
the ``Kepler FOV Field Descriptions'' page on the MAST/KIC web site.
\footnote{http://archive.stsci.edu/search\_fields.php?mission=kepler\_fov}

\section{Photometric Diagnostics and Validation of Stellar Classifications}

The stellar parameters we assigned to stars were, by construction, distributed
in the CMD diagram in plausible ways, and did not conflict with schematic
constraints about the spatial and metallicity distribution of stars in the
galaxy.
But of course these statistical properties of the KIC do not guarantee that
individual stars are correctly classified, even though correct classification
of individual stars is precisely what
the catalog must provide, to serve its purpose.
The classifications can fail at many stages in the process.
It is hopeless to expect good classifications out of bad photometry, but
on many nights the atmospheric conditions were imperfect, or there were
problems with the telescope or other instrumentation.
Thus, we needed diagnostics of the quality of the photometry, with as much
time resolution as we could manage.
Also even good photometry can yield poor classifications if there are
systematic errors in the models to which the photometry is compared, or
if there are true degeneracies in the model-fitting, or if the mathematical
model-fitting problem is ill-posed and unduly sensitive to noise.
For these reasons it was important to test the classification procedures
by comparing their results with independent estimates of the same
stellar parameters, obtained from different data sets.
There proved to be a number of ways to make such comparisons for restricted
sets of stars, and such comparisons have become easier as ground-based
observations have been concentrated on the $Kepler$ field, and as $Kepler$
itself has returned variability information about many target stars.
The degree to which the KIC can be validated nevertheless remains rather
unsatisfactory;  future studies will doubtless give more complete pictures
of the successes and failures of the classification scheme.
In the following sections, we describe the diagnostics that we used as
quality control measures while compiling the KIC, and a few of the validation
tests that we have performed.
 
\subsection{Quality of fits to the photometry}

A simple test of the parameter-fitting process is to compare
observed colors for each star with those predicted for the derived
stellar parameters $\{ \teff, \log(g), \log (Z), E_{B-V}\}$.
We made plots showing this comparison for each 1-degree-square tile
in the {\it Kepler} field.
Within each such tile, we selected
stars for which the fit for stellar properties converged, 
and for which all of the magnitudes $\{ g,r,i,z,D51,J,H,K \}$ had valid
values.
For each star in this set, we
computed model colors.
We then formed the difference between the observed colors and the
model ones.
Each panel of Figure 5 shows the differences plotted as points on a different
color-color diagram (eg $(r-i)$ vs $(g-r)$, or $(J-K)$ vs $(g-i)$).
The colors were chosen so that every filter is represented at least once.
Of interest are the centroids of the plotted clouds of points,
their scatter in each color, and the presence or absence of correlations
between the residuals in different colors.

For the most part, the fitting process appears to have worked well:
for tiles at relatively high galactic latitude (7 degrees and above),
the residuals have nearly zero mean in all colors, with distributions
that are more or less Gaussian.
Some colors have more scatter than others -- this is the case for 
any color involving a 2MASS filter or Sloan $z$; evidently the magnitudes
in these filters are noisier than in the visible-light filters $\{g,r,i\}$ 
by a factor of about 2.
Also, there are clear anticorrelations
between several pairs of colors.
Notably, $(r-i)$ and $(i-z)$ are anti-correlated, as are $(i-z)$ and $(z-J)$,
and also $(J-H)$ and $(H-K)$.
These correlations suggest excess scatter in the common filter of
each of ($i$, $z$, and $H$).
Analysis shows that these anticorrelations arise from two separate
processes, one of which affects exclusively the 2MASS magnitudes, while the
other is exclusively connected with the visible-light data.
Beyond this, the causes of these effects are not yet clear.
Tiles at very low galactic latitudes (less than 7 degrees) typically show
much larger scatter than higher-latitude tiles, and also significant
displacement of the mean residuals from zero.
We suppose that these effects arise from large and spatially
nonuniform interstellar extinction, but as yet we cannot rule out alternative
explanations.
About 3.3\% of the area of the $Kepler$ field of view lies at galactic
latitudes of 7 degrees or less.

\subsection{Giant-dwarf separation in color-color space}

From the perspective of the $Kepler$ mission, the most important
task of the KIC is to correctly distinguish giant stars from dwarfs
across the greatest possible range of $\teff$.
Figure 6 illustrates the photometric basis for 
the KIC's classifications.
Panel (a) shows the
$(g-D51)$ vs $(g-r)$
color-color diagram for one tile covering one square degree of sky 
near $b = 10.5^{\circ}$.
In both panels of Figure 6, stars classified as giants 
(defined as having $\log(g) < 3.6$) are
indicated by heavy red symbols, and dwarfs (with larger $\log (g)$ are
plotted as small black symbols.  
In Figure 6a, main-sequence stars occupy a locus that trends from lower left
to upper right as $\teff$ falls, until $(g-r) \approx 0.65$ is reached.
At this point the $(g-D51)$ colors for dwarfs turn sharply blueward as $(g-r)$
continues to get redder.
At $(g-r) \approx 1.0$ this trend reverses and the $(g-D51)$ color
reddens rapidly.
The result is that main-sequence stars form a sickle shape:
hot stars form the handle of the sickle, and cooler ones form the
(curved, concave upwards) blade.
The area inside the blade of the sickle (and particularly the redward
extension of the handle) is occupied by stars with lower surface gravity,
or with low metallicity.
In these stars the Mg b lines are weak compared to what one sees on the
main sequence, resulting in more flux in the $D51$ band, and a more positive
$(g-D51)$ color.
Most often, plots like these show a clear demarcation between the 
dwarf- and giant-star regions, with few stars having contrary
classifications appearing in either region.

Figure 6b shows
$(J-K)$ plotted against $(g-i)$.
As is well known (eg \citet{bess88}), color-color diagrams
involving $(J-K)$ bifurcate for M-type stars, with main-sequence stars
limited to $(J-K)$ colors smaller than about 0.9,
while low-gravity (and also low metallicity) stars continue to
grow redder with lower $\teff$.
Again, most plots show a clean separation between dwarfs and giants
on this diagram, for $(g-i)$ colors that are red enough.
Note that both panels of the Figure use dereddened colors, where the
reddening $E_{B-V} = A_V/3.1$ is computed from the star's galactic
latitude and estimated distance, as described in Section 7.

\subsection{Comparison of stellar parameters with stars in M67}

Another useful test of the analysis procedure is to compare fitted parameters
for a group of stars with those that can be reliably estimated using
other means.
One group of stars for which this comparison can be done with confidence
is selected from members of the cluster M67.
For this cluster D. Latham and S. Meibom provided a list of 116 stars
that are thought to be single cluster members.
We estimated $\teff$ and $\log(g)$ for these stars from a fit to the
\citet{yi08} isochrone for solar metallicity and an age of 4 GY.
Note that M67 lies far outside the {\it Kepler} field of view, so these stars
do not appear in the KIC proper.
Figure 7a shows the comparison between $\teff$ values for M67 
estimated from the KIC and
those from a stellar evolution model fit to Montgomery's B,V photometry
\citep{mon93}.
With the exception of two extreme outlying stars, the agreement is generally
good.
The $RMS$ difference between the measurements is about 150 K, and the systematic
differences appear to be small.
Figure 7b shows the analogous comparison for $\log (g)$.
Again the general agreement is good, with an RMS difference between
the isochrone measurements and the KIC fits of about 0.4 dex.
Systematic differences are discernible in $\log (g)$, however.
The most significant of these is a tendency for subgiants identified via
isochrone fitting
appear as main-sequence stars in the KIC analysis.
This is not surprising, since in the $\teff$ range where the turnoff to
the subgiant branch occurs in M67, none of the filter combinations that
were observed for the KIC are sensitive to gravity.
The color $(g-D51)$ shows a useful gravity dependence only for
$(g-r)$ colors greater than 0.65, whereas the main-sequence turnoff in
M67 lies at bluer colors, roughly $(g-r) = 0.38$.

\subsection{Comparisons with spectroscopic parameter estimates}

We and others have made comparisons between KIC estimates of stellar parameters
and various sets of parameters estimated from modeling of optical spectra.

\citet{mol2010} obtained spectra for 109 relatively bright KIC stars, 
spanning a wide range
of $\teff$.
They found that for temperatures below 7000 K, KIC $\teff$ values agree with
their spectroscopic estimates within about $\pm 200$ K, but that at higher
temperatures, larger deviations occur.
The largest errors appear for the hottest stars;  indeed, in this sample
there are 9 stars with spectroscopic $\teff$ values in the range 9000 K
to 13500 K, and all of these stars are shown in the KIC with lower temperatures,
with differences as large as 4000 K.
As mentioned above, these failures of temperature estimation are expected
for hot stars, because of the absence of $u$-band data.
\citet{mol2010} found that the KIC surface gravity estimates were fairly
accurate for dwarfs, but for giants (including those with $\log (g)$ as
low as 1.5), the KIC estimates could be in error, sometimes by as much
as 1.5 dex.

In the process of studying candidate planet host stars, the $Kepler$ mission
has obtained high signal-to-noise Keck/HIRES spectra of a few tens of stars, 
and D. Fischer
has analyzed these with the Spectroscopy Made Easy (SME) package 
\citep{val96} to estimate
their values of $\teff$, $\log (g)$, and [Fe/H].
Figure 8a shows the comparison for all 3 parameters for 34 of these stars
(all that were available as of Sep. 2010).
The selection criteria for these {\it Kepler} planet candidates assured that
this sample of stars consists almost entirely
of dwarfs with roughly solar $\teff$.
Thus, the $\teff$ range is considerably smaller than for the sample
measured by \citet{mol2010}.
For all but 2 of these stars, the difference between KIC and HIRES/SME
values of $\teff$ is 200 K or less; the RMS difference is 135 K.
There is some evidence for a systematic trend in the $\teff$ differences,
with the KIC temperatures being cooler then Keck/HIRES at 
low $\teff$ and warmer at high
$\teff$, but more measurements are needed to confirm this impression. 
Figure 8b compares the KIC and HIRES/SME estimates of $\log (g)$.
Except for one star (spectroscopically classified as a subgiant with
$\log(g)$ = 3.5), the two sets of estimates agree within $\pm$0.3 dex,
and the RMS difference 
is only 0.25 dex.
We consider that this agreement is largely artificial, however. 
The $Kepler$ planetary transit candidates consist almost entirely of stars
classified as dwarfs in the KIC, and hence included in the {\it Kepler} target list.
If these stars are observed to show
photometric transit events, then the original classification is likely correct.
Thus, for stars selected as these were, we expect at least rough agreement 
concerning $\log(g)$.
Figure 8c compares the KIC and HIRES/SME values of $[Z]$ for the same stars
(excluding a few stars in the previous plots for which HIRES/SME values 
of $[Fe/H]$ were not reported).
The total range of $[Fe/H]$ estimated by HIRES/SME for these stars is small,
about 0.7 dex, and most of the stars are clustered in about half of this
range.
Accepting (for lack of an alternative) the shortcomings of this sample,
one finds that the RMS difference between the two sets of estimates is
0.2 dex, and that the Spearman rank correlation coefficient is 0.42, with
a two-sided significance of its deviation from zero of 0.02. 
For comparison,
the Spearman statistics for the different $\teff$ measurements are 0.96
and $5 \times 10^{-17}$, respectively.
Thus, while there is evidence that the KIC values of $[Z]$ are 
related to those measured spectroscopically,
the strength of this relationship is unimpressive.
Moreover, there appears to be a significant systematic difference
between the KIC and HIRES/SME values, in the sense that the KIC $[Z]$
values are about 0.17 dex smaller.
We suspect this is symptomatic of the Bayesian prior for $[Z]$ being
rather narrowly peaked around $[Z] = -0.1$, whereas planet-bearing
stars (which are abundant in this sample) are typically metal-rich 
(e.g, \citet{fis05}).
 
The $Kepler$ mission has also observed a much larger sample of stars
using relatively low signal-to-noise ratio spectroscopy 
(S/N of typically 7 to 10) 
obtained from several different
sources (McDonald 2.7m, Mt. Hopkins 1.5m, Lick 3m, and Nordic Optical
Telescope 2.5m telescopes).
These spectra were obtained to facilitate
identification of stellar binaries and to provide crude spectral 
classifications, 
so as to make an early decision
about the likely origin of apparent photometric transit signals.
These ``reconnaissance'' spectra were analyzed by correlating them against
templates created from stellar atmosphere models, using a grid spacing
of 250 K in $\teff$ and 0.5 dex in $\log (g)$, and assuming solar
metallicity.
The stars in this sample were commonly observed spectroscopically
2 or 3 times each.
Again, by virtue of being selected as transiting planet candidates,
these stars form a biased sample, favoring dwarfs.
In Figure 9 we plot the average (over observations) of the $\teff$
and $\log (g)$ values estimated for each star against the similar values
found in the KIC. 

The agreement between KIC classifications and the ones from reconnaissance
spectroscopy are worse than for the KECK/HIRES/SME classifications, with
random $\teff$ differences of about {360 K} RMS, random $\log(g)$
differences of roughly 0.3 dex, and evidence for systematic
errors of similar magnitude.
On the other hand, comparisons between successive spectroscopic estimates of 
$\teff$ and $\log(g)$ for any given star show scatter of similar size.
Thus, a substantial fraction of the scatter in Figure 9 likely results from
errors in the spectroscopic reconnaissance measurements.
Improved analysis techniques for reconnaissance spectroscopy will soon
provide better material for assessing errors in the KIC.
In the meantime, we find that KIC $\teff$ and $\log(g)$ estimates agree with
those from reconnaissance spectroscopy about as well as the latter agree with
each other.
  
\subsection{Comparisons with asteroseismic parameter estimates}

Many stars with KIC classifications have been observed to oscillate in global
modes, usually p-modes.
Indeed, early {\it Kepler} observations are the source of the vast majority of
these pulsation detections.

In a simple test, \citet{koc2010} compared the RMS photometric 
variability of 1000 stars with $\teff \leq 5400$ K,
that the KIC classifies as giants, with an equal number classified
as dwarfs.
Giants are known to be systematically more variable, because virtually
all of them show p-mode oscillations with amplitudes that increase with
increasing stellar luminosity.
In these samples, about 2.5\% of the stars classified as dwarfs showed
variability consistent with their being giants, 
and none of those classified as giants had
variability consistent with dwarfs.
It thus appears that, averaged over this sample, the KIC is more than 98\% 
successful in its principal aim --  to
distinguish between cool giant and dwarf stars.

Detailed seismic analyses have been published for a few Sun-like dwarfs
in the {\it Kepler} field.
\citet{jcd2010} used {\it Kepler} time series to search for p-mode frequencies 
and estimate stellar parameters in 3 {\it Kepler}-field stars that were known
from groundbased observations to host transiting planets.
All of these stars (HAT-P-7, HAT-P-11, and TrES-2) are however too bright
to avoid saturation in the SCP photometry, hence have no $\teff$ or $\log(g)$
values in the KIC.
\citet{cha2010} analyzed {\it Kepler} time series for 3 fainter Sun-like stars,
namely KIC {6603624, 3656476, 11026764}.
The analysis included groundbased high-resolution spectroscopy (allowing
estimates of $\teff$, $\log(g)$, and $\log(Z)$), and p-mode fitting, which
when combined with the non-seismic data, allowed estimates
of the stellar mass and radius.
For these stars, the KIC estimates of $\teff$ were all lower than the
spectroscopic ones, by \{-374 K, -242 K, -138 K\} respectively. 
The KIC estimates of $\log(g)$ were all larger than the seismic estimates,
by \{0.064, 0.253, 0.066\} dex, respectively.
\citet{met10} did an independent analysis of KIC 11026764, finding
$\teff$ and $\log(g)$ values that are consistent with those by \citet{cha2010}.

Early {\it Kepler} data have revealed long-lived p-modes in a large number
of giant stars.  A recent study by \citet{hek2011} has made an explicit
comparison between KIC and asteroseismic estimates of $\log(g)$ for
a sample of 11805 stars classified as giants in the KIC that also have
Quarter-0 and Quarter-1 {\it Kepler} time series available to the public, and in which
p-modes have been detected.
These authors used the method described by \citet{kal2010} to estimate
stellar masses and radii from the p-mode large frequency separation
$\Delta\nu$, the frequency of maximum power $\nu_{max}$, and the
KIC estimate of $\teff$.
This comparison shows that while the KIC correctly ascribes low surface
gravities ($\log(g) \leq 3.8$) to almost all of these stars, the KIC values
are systematically too large relative to the asteroseismic ones.
The magnitude of this error is larger for lower gravities; for
clump giants, with $2.3 \leq \log(g) \leq 2.7$, it is about 0.5 dex.
The KIC values also show larger scatter at given $\teff$ than do the 
asteroseismic ones.

\vskip12pt
\subsection{Comparison with $Hipparcos$ Parallaxes}

For a small sample of stars, one may usefully compare the distances inferred
from the KIC analysis with parallaxes measured by the $Hipparcos$ mission
\citep{hog00}.
As described in \S 7, the distance $d$ to an observed star is implied by
its apparent $r$ magnitude, its galactic latitude, and the exponential
model that we use for interstellar extinction.
The distance $d$ is not tabulated in the KIC, but it may be computed from
the (tabulated) $V$-band extinction $A_V$ as
$$
d \ = \ -{h \over \sin b} \ln (1 \ - \ A_V {\sin b \over {k_V h}}) \ , \eqno(20)
$$
where $h$ is the dust scale height (assumed to be 150 pc), $b$ is the 
galactic latitude, and
$k_V$ is the dust opacity in the galactic plane (assumed to be 1 mag pc$^{-1}$).

Unfortunately, stars with reliable ($\pm$ 20\%) $Hipparcos$ distances tend
to be brighter than typical stars with valid (not saturated) SCP photometry.
Thus, the number of stars having both 20\%-accurate $Hipparcos$
parallaxes and converged
parameter solutions is only 55, and these lie at distances that are
considerably smaller than those of typical $Kepler$ stars.
Figure 10 shows the comparison of parallax $vs$ extinction distances for
this sample of stars.
These stars span magnitudes $7.88 \leq r \leq 10.84$, with 
colors $-0.47 \leq g-r \leq 0.87$, and KIC temperatures 4684 K $\leq \teff \leq$
10735 K.
The absolute magnitudes we compute for these stars (based on the parallaxes)
show that all but 2 lie on the main sequence.
About 65\% of these stars have parallax- and 
extinction-based
distances that agree within a factor of 1.7, or $\pm 0.23$ dex.
Little of this scatter can be attributed to the parallaxes,
so this comparison suggests that KIC stellar radii have 1-$\sigma$ errors
that are also about 0.2 dex.
This is roughly consistent with our previous estimate of 0.4 dex for the
typical error in $\log(g)$.

Figure 10 also suggests that (aside from 2 dramatic outliers) the extinction 
distances are systematically too small.
For this sample of stars, the median distance error is -0.12 dex, or a
factor of 0.76.
There are two possible explanations for this bias:  the in-plane dust opacity
$k_V$ may be overestimated, or the luminosities that we attribute to dwarf
stars may be too small.
We believe that too-small luminosities are mostly at fault, since this
explanation is consistent with our previous conclusion that we tend
to overestimate $\log(g)$ for dwarfs.

There are also 2 stars for which the extinction distances are much too large.
One of these is a binary G-type dwarf that the KIC evidently misclassified
as a giant, giving a distance that is much too large and $\log(g)$ that
is much too small.
The other, HIP93941 = BD+42$^{\circ}$3250, 
is classified by \citet{cat10} as a B2 star with
weak He lines,
and by \citet{ost10} as an sdB star.
Thus, its temperature is surely much hotter than the KIC value of 10735 K,
and its parallax yields $V_{abs} = 4.4$.
This stellar class is not
represented in the Bayesian prior distribution used in the KIC analysis,
and hence the object was misclassified as a more luminous and distant star
than it actually is.

\vskip12pt
\section{KIC Shortcomings}

As indicated above,
the stellar classifications provided in the KIC suffer from several known
systematic defects that should be considered when using the catalog.
Here we describe (or repeat) the most important of these, explain their
source when this is known, and illustrate the problems with samples from
the data.

\vskip12pt
\subsection{$\teff$ Scale}

KIC $\teff$ values have systematic disagreements with other $\teff$
estimates that apply to the same stars.
(Of course, these other estimates also disagree systematically with each other.)
For approximately Sun-like stars these disagreements are usually 
less than 50K, 
though in the worst cases
they may exceed 200K.
For stars that are distant from the Sun on the CMD, one must
be more cautious.
The KIC $\teff$ estimates are untrustworthy for $\teff \ge 10^4$K, and
also for $\teff \le 3750$K.

For hot stars ($\teff \ge 9000$ K), the lack of $u$-band data makes our
photometry insensitive to variations in $\teff$.
Higher temperatures are found in the KIC estimates, but their values should
not be trusted.
We used a subset of the CK models with a maximum $\teff$ of 19,000K;
the absence of estimates above this value therefore does not imply the
absence of such stars in the sample.

The CK model atmospheres we used covered only $\teff \ge 3500$K,
and we applied ad hoc corrections to the colors for $\teff \le 3750$K.
Temperatures below the latter value are therefore also questionable
(although at fixed composition and gravity, the KIC $\teff$ is probably at
least a monotonic function of the true $\teff$).

\vskip12pt
\subsection{Subgiant gravity}

The KIC classifications tend to give $\log (g)$ too large for subgiant
stars, especially those hotter than about 5400K.
This leads to underestimates of the radii of this subset of stars, typically by factors of
1.5 to 2.

For temperatures above roughly 5400K, none of our photometric colors provide
information about surface gravity.
Accordingly, for hotter temperatures the maximum posterior probability analysis 
has no basis
to choose any $\log (g)$ other than the one that is most probable a priori,
which corresponds to the center of the main sequence, near $\log (g)=4.5$.
Stars on the giant branch almost all have $\teff$ small enough so that their
gravities can be measured;
so the gravities of true giants appear to be estimated with errors of typically
about 0.5 dex.
But for
a significant subset of (mostly hot) stars, the KIC-derived gravities
are systematically too large, sometimes by more than 1 dex.

Given the available photometric data, this problem is essentially unavoidable.
The information needed to distinguish between F- and early G-type dwarfs and
subgiants is not present in the photometry, and there is no way to obtain sensible
results without it.
(Biasing the prior probabilities towards lower gravities, for instance, results
in more subgiants, but with no guarantee that the new alleged
subgiants are in fact the stars with low gravity.)
Users should thus be wary of $\log (g)$ estimates for $(g-r) \leq 0.65$.

\vskip12pt
\subsection{High $\log (Z)$ at low $\teff$}

We could perform few tests of the veracity of the estimates of $\log (Z)$,
and (given the absence of $u$-band magnitudes) there is little reason to
trust these estimates.
The $(g-D51)$ color contains information about $[Z]$, but this is almost
entirely degenerate with the larger and more common color perturbation
caused by surface gravity.

A plot of $\log(Z)$ vs $\log (\teff)$ for a large randomly-sampled group
of stars (Figure 10) shows a number of peculiarities.
For $\teff$ below about 4200K ($\log (\teff) = 3.623$), 
the estimated $\log(Z)$ distribution begins
to spread and bifurcate, and below 3800K ($\log (\teff) = 3.580$), 
virtually all stars show $\log(Z)$
greater than +0.5, which is the highest metallicity represented in the
CK models that we use.
This behavior has not been investigated in detail, but it seems likely that
it results from a mismatch between the model and observed color dependences
at low temperatures, in the sense that (other astrophysical evidence
notwithstanding), high-metallicity stars provide the best fit to the observations.

One can also observe clustering of $\log(Z)$ around integral and half-integral
values of $\log (Z)$, for $\log (Z) \le -1$.
These are the tabulated values of $\log (Z)$;  the concentration of estimated
values near the tabulated ones presumably indicates a failure of the
interpolation and fitting code that optimizes the posterior probability.

The most encouraging demonstration that the estimates are doing something
sensible was provided by the classifications of stars in the globular clusters
M13 and M92.
These clusters showed a fairly large (but by no means dominant) fraction of
low-$Z$ star classifications.
Based on their positions in our observed cluster color-magnitude diagrams,
all of these cluster stars were, however, cool giants, for which we suspect
the model colors are particularly uncertain.
Thus, whether the classifications for metallicity are performing properly 
for main-sequence stars
is unknown at present;  it would be prudent to assume that they are not.
Fortunately, the fraction of low-metallicity stars in the solar neighborhood is
quite small, so for the purposes of the {\it Kepler Mission} the uncertainty about
$\log (Z)$ is tolerable.
But anyone with a particular interest in stellar metallicities should not use
the KIC for their estimates of $\log (Z)$.

\vskip12pt
\subsection{Extinction and Reddening}

Regions at low galactic latitude are prone to have large and spatially
nonuniform extinction and reddening.
The model of extinction that is employed in the Bayesian posterior probability
maximization contains no small-scale structure, so it is unable to deal
properly with localized large deviations from the mean extinction.
the result is systematic misclassification of many stars, a scattered
and confused relation between $\teff$ and color, and other failings.
Examples of such difficulties are shown in Figure 11, 
which compares tiles at low and high galactic latitudes.

\section{Summary and Conclusions}

The {\it Kepler} Input Catalog is available via the MAST archive facility, operated
by the Space Telescope Science Institute \footnote{http://archive.stsci.edu/kepler/kic10/search.php}.

Experience with the KIC, combined with the testing that we report here and
that has been done elsewhere, shows that the KIC has succeeded in its
primary goal --  to distinguish between cool giant and dwarf stars with
good reliability, so that the {\it Kepler Mission} can select optimum
targets for its transiting-planet search.
As a by-product of that goal, the KIC provides 
photometry in the SDSS-like photometric bands $g,r,i,z$ and in the
intermediate-bandwidth $D51$ bandpass, calibrated with a typical 
flux accuracy of about 2\%, for stars brighter than about magnitude 14 in any 
of our filters.
All of this information is federated with that from other key photometric
and astrometric databases, so that the KIC can serve as a tool for
research on a great number of objects that will not be observed by
$Kepler$ itself.

Experience and testing has also shown that the KIC has defects.  The most
notable of these
include stars that appear in other catalogs but that have no physical
classifications in the KIC, 
systematic errors in estimates of $\teff$ for hot and for very cool stars,
systematic errors in estimates of $\log(g)$ for stars with $g-r$ colors
that are bluer than about 0.65,
and questionable metallicity determinations across the CMD.
Most of these problems arise from a common cause, namely lack of information
about the desired physics in the mostly wideband photometry that we
were able to obtain.
By combining $u$-band, and perhaps also suitable intermediate-bandwidth 
observations with
the techniques described here, it should be possible to extend greatly
the $\teff$ range over which the KIC parameter estimates are reliable,
and to improve substantially the KIC's metallicity sensitivity.
Also, careful spectroscopic observations of stars that have KIC
classifications should allow better characterization of the KIC's
systematic errors.
We hope that others will find it useful to provide these improvements.

We are grateful to the Mt. Hopkins support and observing staff, especially
Carl Hergenrother, for his tireless work obtaining the necessary
observations.  
We also thank Dave Monet, for his indispensable help in implementing the
KIC astrometry methods, and for federating the SCP with other catalogs
that carry essential information.
We thank Steve Howell, for helping to define our photometric
approach in the project's early days.
We thank Geoff Marcy, Howard Isaacson, Debra Fischer,
Bill Cochran, Sam Quinn, Lars Buchhave, Mike Endl, and Phillip MacQueen
for the use of
their spectra and spectral analysis of $Kepler$ target stars.  We also
thank Saskia Hekker, Bill Chaplin, Ronald Gilliland, and dozens of members
of the {\it Kepler} Asteroseismic Consortium for making their seismic data
available before publication.  We are deeply grateful to the $Kepler$
Science Team and everyone connected with the {\it Kepler Mission},
for keeping the mission running smoothly, and for providing the amazing
$Kepler$ photometric data, the promise of which was the inspiration for 
the current work.  
We thank Jeffrey Scargle for a careful reading of an early version of
this work, and
Don Kolinski for extensive software development help.
T.B. acknowledges support from NASA Grant Number NNX10AG02A, and
thanks HAO/NCAR and Las Cumbres
Observatory Global Telescope for patience and support while this work 
was being done.  The National Center for Atmospheric Research is supported
by the National Science Foundation.
We are grateful to the {\it Kepler Mission} for partial support of the
photometric observations under NASA Cooperative Agreement NCC-1390
with the Smithsonian Astrophysical Observatory.

\clearpage

\centerline{\bf Figure Captions}

\figcaption[MtHop_zero_pt.eps]{The
combined atmospheric and instrumental coefficients $a_i$
(see Eqn 1)
for the $g$, $D51$, $r$, $i$, and $z$ filters for
203 of the 205 nights on which KIC data were obtained. The remaining
2 nights gave values that are extreme outliers, falling outside the
range of these plots.
Zero points on these curves have been shifted for plotting convenience.
Note the color dependence of the temporal variations,
which tend to be larger in blue filters than in red.
Vertical dashed lines indicate the dates of transition between CCD cameras.
\label{Figure 1}}

\figcaption[New_Fig4.ps]
{Plot of $g$ extinction data on a representative
(mildly non-photometric) night.
The top panel shows extinction as a function of airmass.
Points plotted as diamonds were obtained before the meridian transit of
the standard star field;  thost plotted as triangles were obtained after it.
Vertical bars indicate the interquartile spacing of results from the 20 stars used to
estimate the extinction.
The diagonal dashed line is the result of a robust linear fit to the extinction
values, with coefficients tabulated in the upper left corner of the plot.
The bottom panel shows residuals around this fit plotted against the time-varying
FWHM of the stellar point spread function.
\label{Figure 2}}
 
\figcaption[refvsx_k_coeff.eps]{Variation in the $k_i$ coefficient 
(extinction per
unit airmass in the $i^{th}$ filter), shown as a function of date, for
nights for which reliable linear fits to the extinction could be obtained,
during a part of one observing season.
The time span shown here was one of the most variable that we encountered
in 5 seasons of observing, but is nonetheless typical in its variability
within a factor of 1.5. 
Filters $g,D51,r,i,z$ are shown in order from top to bottom of the plot.
Night-to-night variations tend to be not only correlated, but of similar
size among the filters.
An exception is the $z$ filter, in which the extinction is evidently affected
by a different process than at shorter wavelengths.
We believe that this process is extinction from water vapor.
\label{Figure 3}}

\figcaption[rms_vs_r.eps]{Photometric repeatability for stars in
the secondary standard fields, after correction for
atmospheric extinction for each filter, shown as a function of apparent
magnitude in that filter.
\label{Figure 4}}

\figcaption[Fig14.ps]
{Comparison between observed and model colors.
This plot represents all of the stars contained in one tile,
covering an area spanning 1 degree in RA by 1 degree in Dec on the sky
near galactic latitude $b \simeq 10.5^{\circ}$.
Each point corresponds to one star, and the plotted positions show the
residuals (in magnitudes) after subtracting the best-fit model from the
photometry for that star, for two chosen colors (eg $g-i$ vs $g-r$, as
in panel a).
Different panels show various combinations of colors, indicated in
the axis labels.
The tile plotted here (at RA=292$^{\circ}$ and Dec =+40$^{\circ}$) 
is fairly typical of areas in which the
model fits are good.
There are several notable features.
In panels a and b,
the $RMS$ scatter in the residuals is .02 mag or less in each of $g-r$,
$r-i$, and $g-D51$.
Errors are anticorrelated between $g-r$ and $g-i$, but this tendency is more
noticeable and has a different slope in the wings of the distributions than
in the cores.
In panels c and d,
note the larger scatter (especially in $z-J$,
and also the strong negative correlations between these pairs of
residuals.
In panels e and f,
note the change in plot scale;  the residuals in the IR colors are
much larger than in the visible bands.
The $J-K$ and $g-i$ residuals are almost uncorrelated, but one now sees
a very significant offset from zero of the mean residual in $g-i$.
The two IR colors in the bottom panel have only slightly correlated
residuals, but the center of the $J-H$ distribution is also far
displaced from zero.
\label{Figure 5}}

\figcaption[giant_v_dwarf.eps]
{Panel (a) shows the $(g-D51)$ vs $(g-r)$
color-color diagram for the same 1-degree-square tile as shown in the
previous Figure,
at galactic latitude $b \simeq 10.5^{\circ}$.
Stars classified as giants (with $\log g) < 3.6$) are 
indicated by red symbols, and dwarfs (with larger $\log (g))$ are
plotted in black.
Panel (b) is similar to the above, but shows a $(J-K)$ vs $(g-i)$
color-color diagram for the same tile.
The meaning of the symbols is the same.
\label{Figure 6}}

\figcaption[New_Fig11.eps]{Stellar parameter estimates for confirmed single 
stars in the star cluster
M67.
Estimates from the KIC are plotted on the $y$-axis;
along the $x$-axis are values from the
4 GY solar-abundance Yonsei-Yale isochrone \citep{yi08},
fit to B,V photometry by \citet{mon93}.
The diagonal dashed line indicates equality.
Panel (a) shows $\teff$ on each axis, with lines showing equality and $\pm$200 K
overplotted.
Panel (b) similarly shows $\log(g)$, with lines showing equality and
$\pm$0.3 dex.
\label{Figure 7}}

\figcaption[HIRES_parms.eps]{Parameter estimates from the KIC 
plotted against estimates
by D. Fischer from SME model spectrum fitting, for stars observed with
the Keck/HIRES spectrometer as possible transiting planet hosts.
A few stars have more than one independent HIRES/SME observation and
analysis;  in these cases the various results are shown connected by
horizontal lines.
Panel (a) shows $\teff$ on each axis, with lines showing equality and $\pm$200 K
overplotted.
Panel (b) shows $\log(g)$, with lines showing equality and
$\pm$0.3 dex.
Panel (c) similarly shows $\log(Z)$, with line showing equality and 
$\pm$0.4 dex.
\label{Figure 8}}

\figcaption[Recon_parms.eps]{Same as the previous Figure, except for stars with $\teff$
and $\log(g)$ estimates from matching a grid of models to reconnaissance 
spectroscopy of {\it Kepler} transiting planet candidates.
Most plotted points are the average of results from 2 or more 
independent spectra.
Also, small random offsets have been applied in both axes, to reduce crowding.
Diagonal lines indicate equality and $\pm$250 K, $\pm$0.3 dex.
The analysis of these spectra provided no metallicity estimates, so $\log(Z)$
is not shown.
\label{Figure 9}}

\figcaption[hipp_kic.eps]{Comparison between KIC ``Extinction'' distances 
and $Hipparcos$ ``Parallax'' distances to stars in the Kepler field,
having valid KIC stellar parameters and parallaxes accurate to 20\% or better.
Error bars correspond to $\pm$ 1-$\sigma$ uncertainties on the parallaxes.
Diagonal lines indicate equality and $\pm$ 0.23 dex (factor of 1.7).
\label{Figure 10}} 

\figcaption[FigA4.ps]{Plot of KIC estimates of 
$\log (Z)$ against $\log(\teff)$, illustrating
the strong tendency of low-temperature stars to be classified with high
metallicity, and moreover the tendency for low-$Z$ stars to be ascribed
integral- or half-integral values of $\log (Z)$, independent of $\teff$. 
\label{Figure 11}}

\figcaption[color_vs_teff.eps]{The relation for $(g-r)$ color (uncorrected
for interstellar reddening) and $\log(\teff)$, for a tile near the galactic
plane, at $b \simeq 6^{\circ}$ (panel a), and at $b \simeq 18^{\circ}$
(panel b).
Besides having more stars per unit area near the plane, the near-plane
scatter diagram shows general reddening, increased spread in color at
each $\teff$, and a complex, roughly bimodal structure.
\label{Figure 12}}

\clearpage\par

\begin{figure}
\plotone{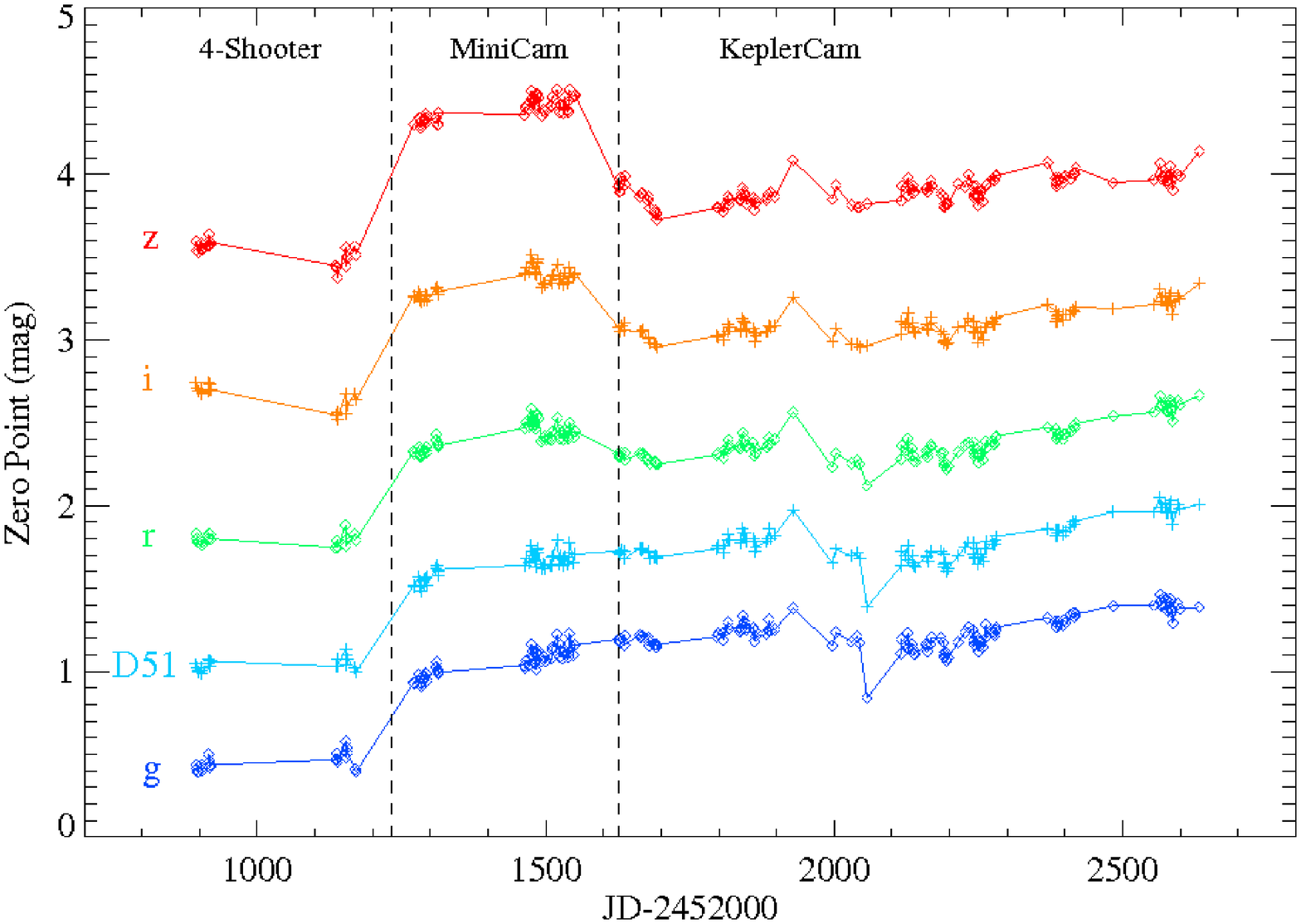}
\end{figure}

\clearpage

\begin{figure}
\plotone{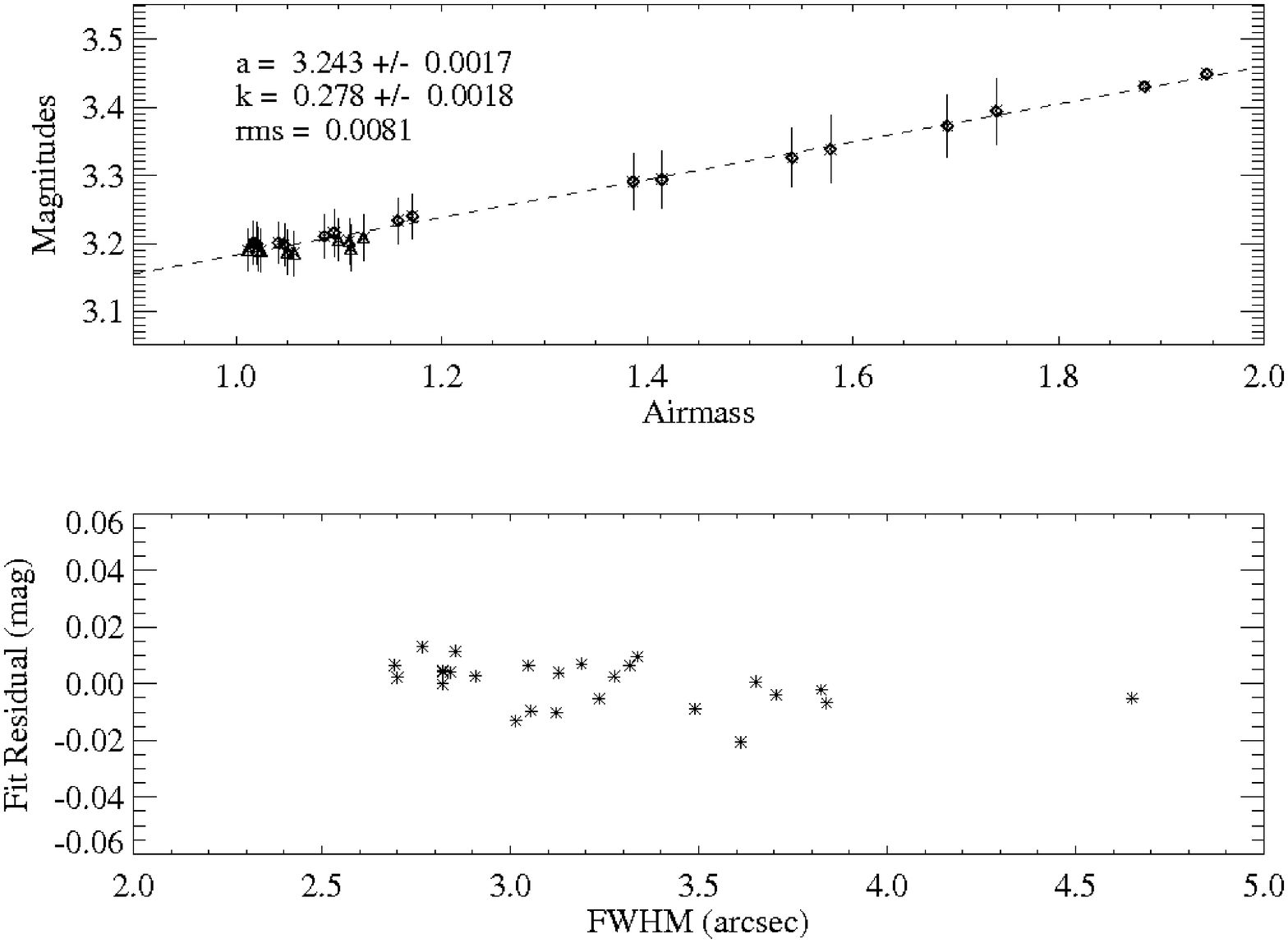}
\end{figure}

\clearpage

\begin{figure}
\plotone{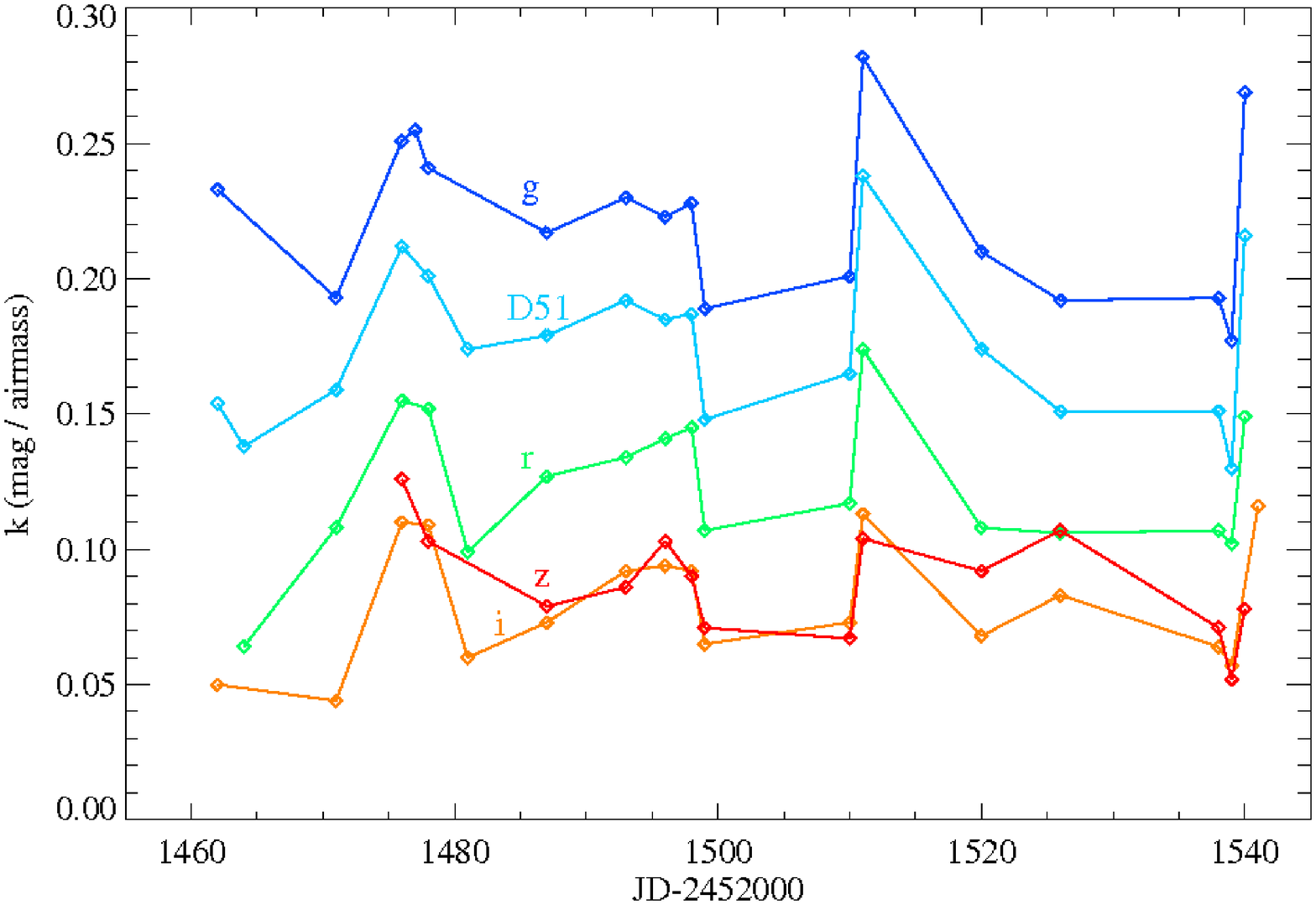}
\end{figure}

\clearpage

\begin{figure}
\plotone{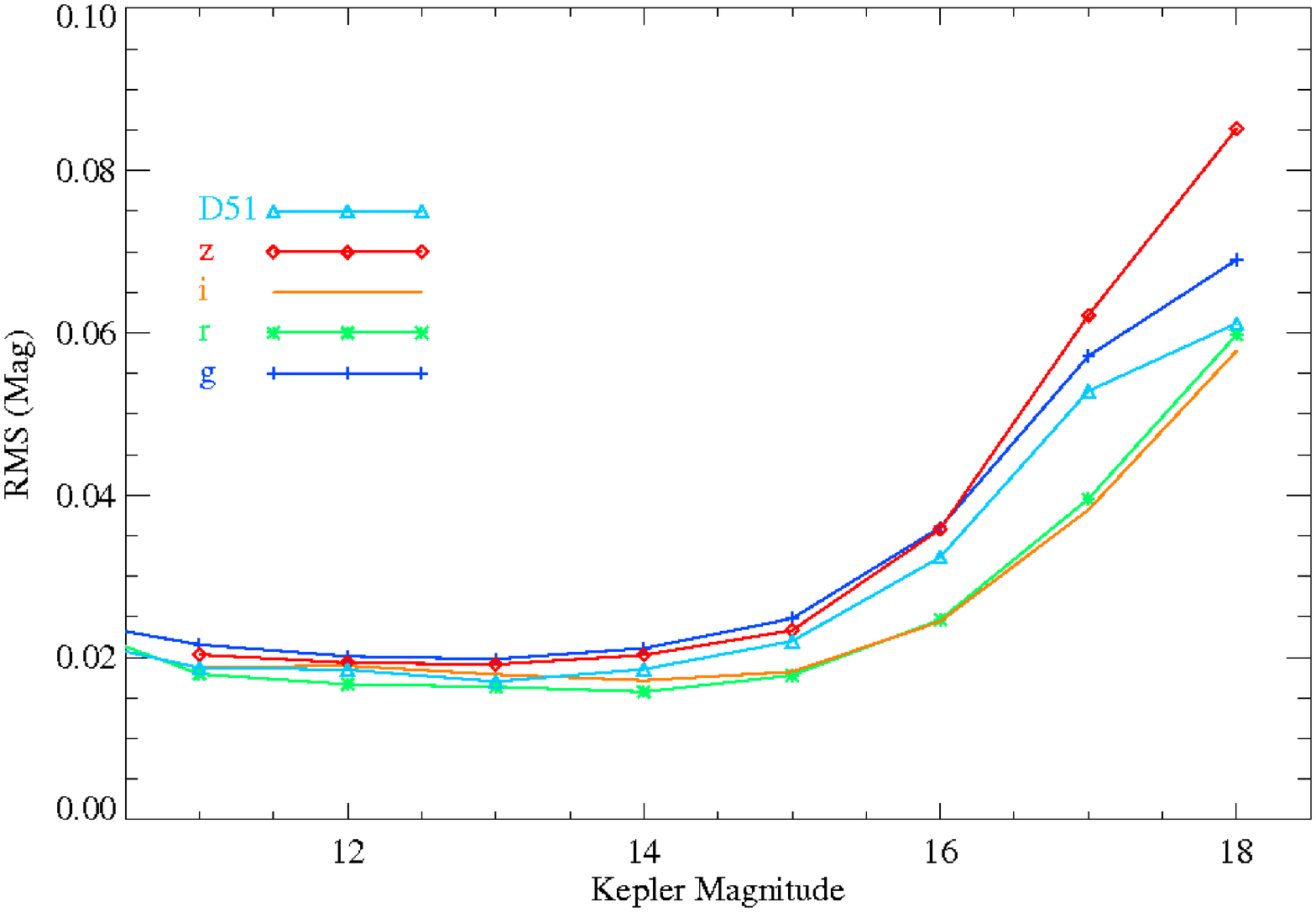}
\end{figure}

\clearpage

\begin{figure}
\plotone{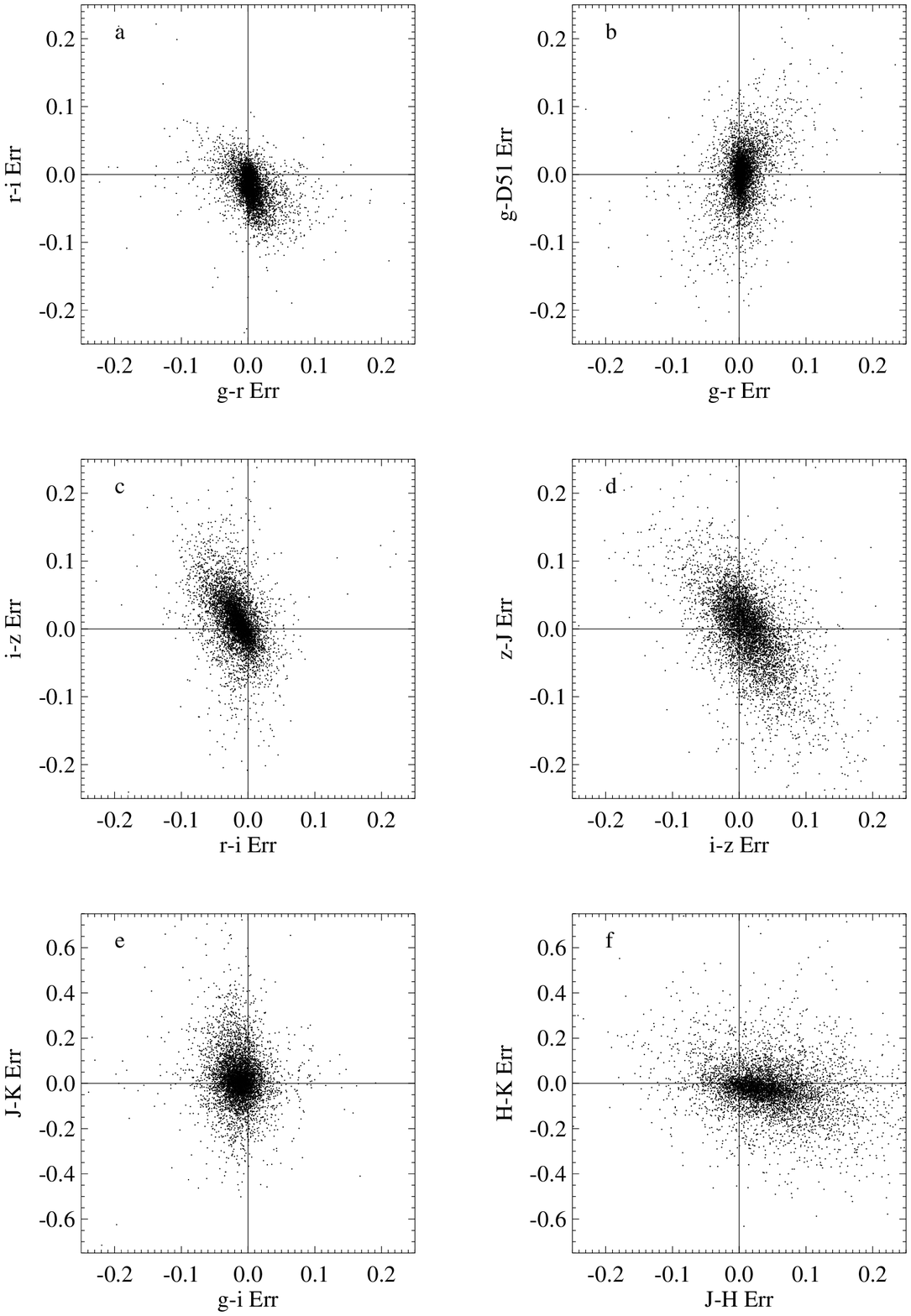}
\end{figure}

\clearpage

\begin{figure}
\plotone{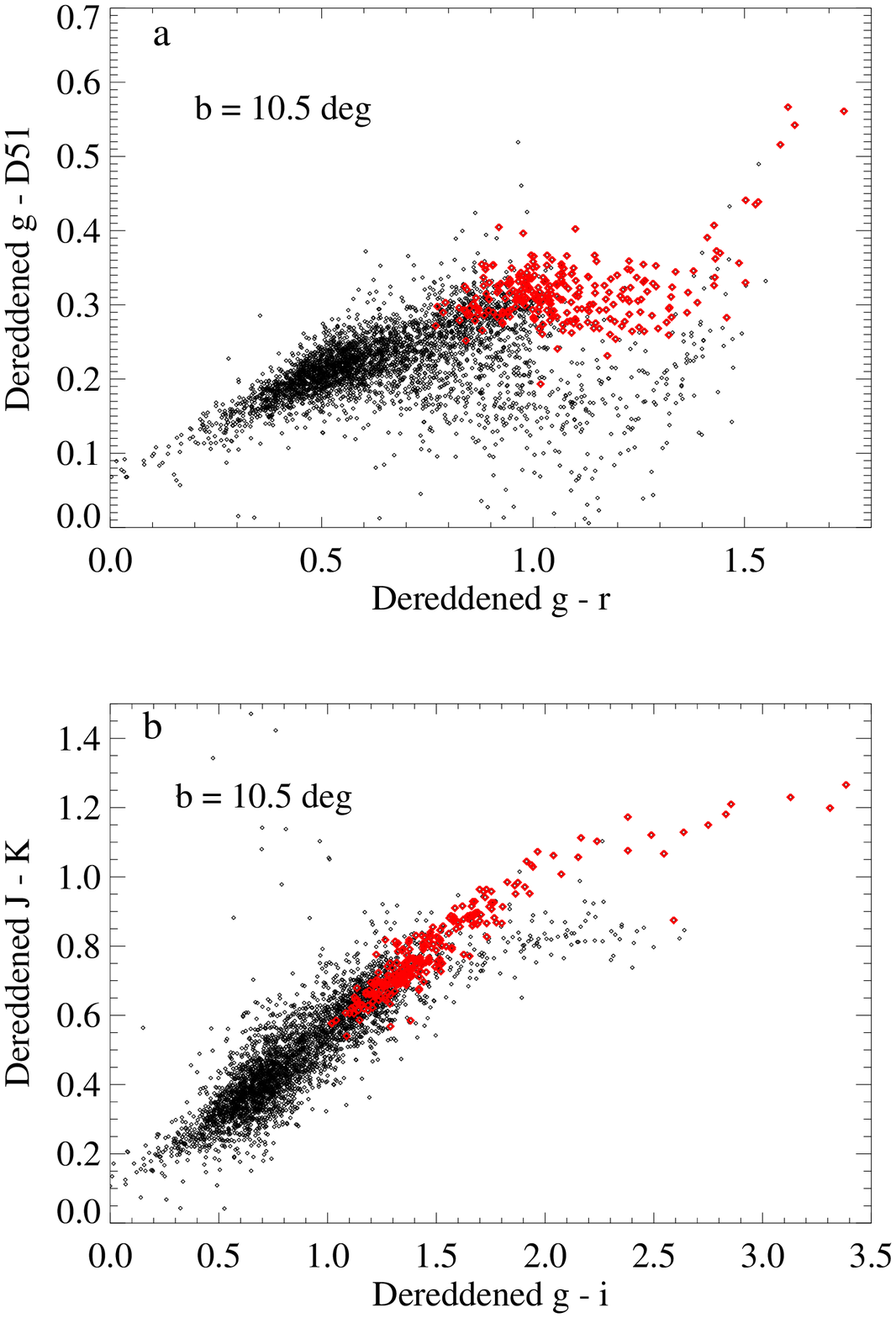}
\end{figure}

\clearpage

\begin{figure}
\plotone{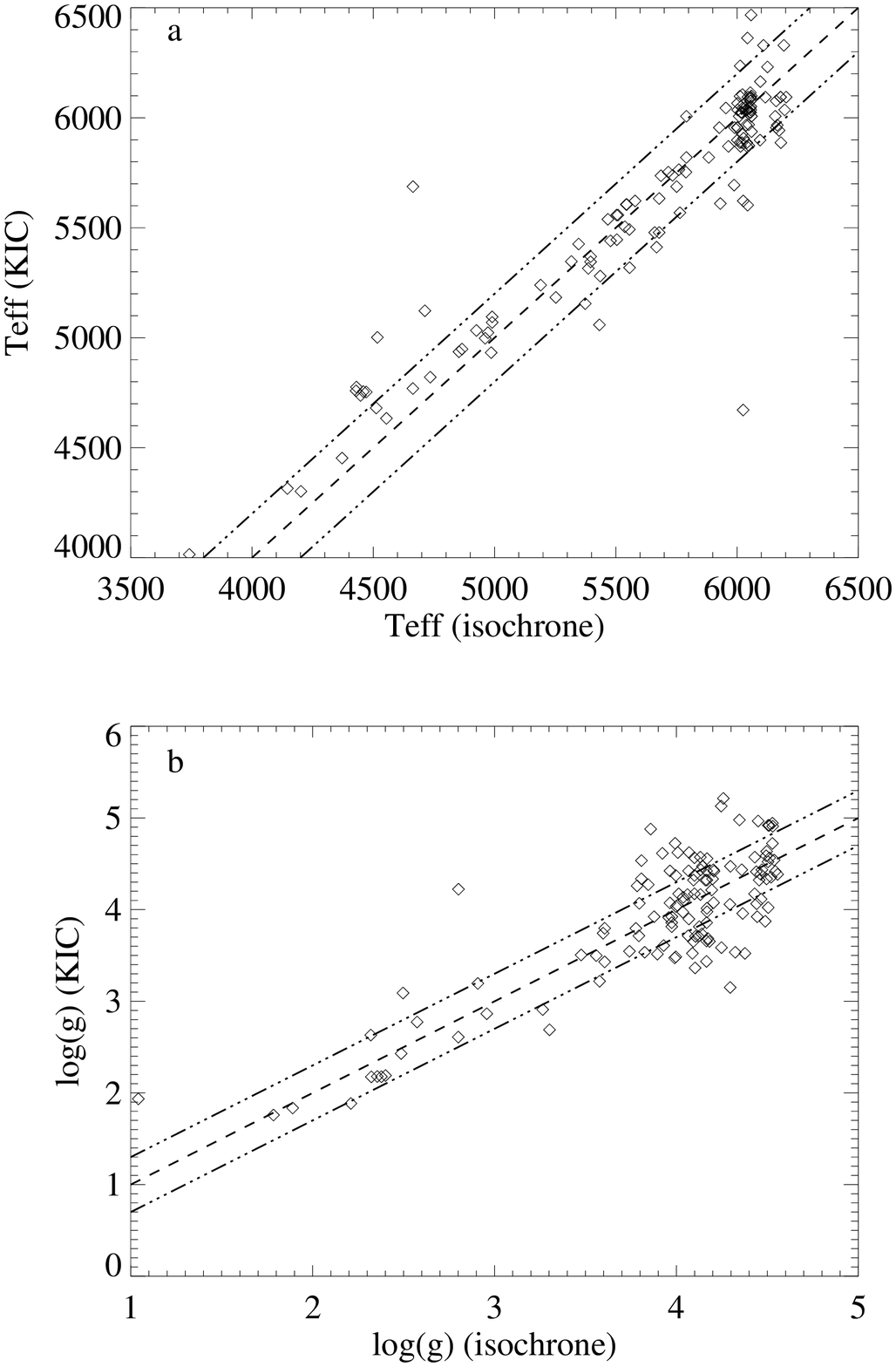}
\end{figure}

\clearpage

\begin{figure}
\plotone{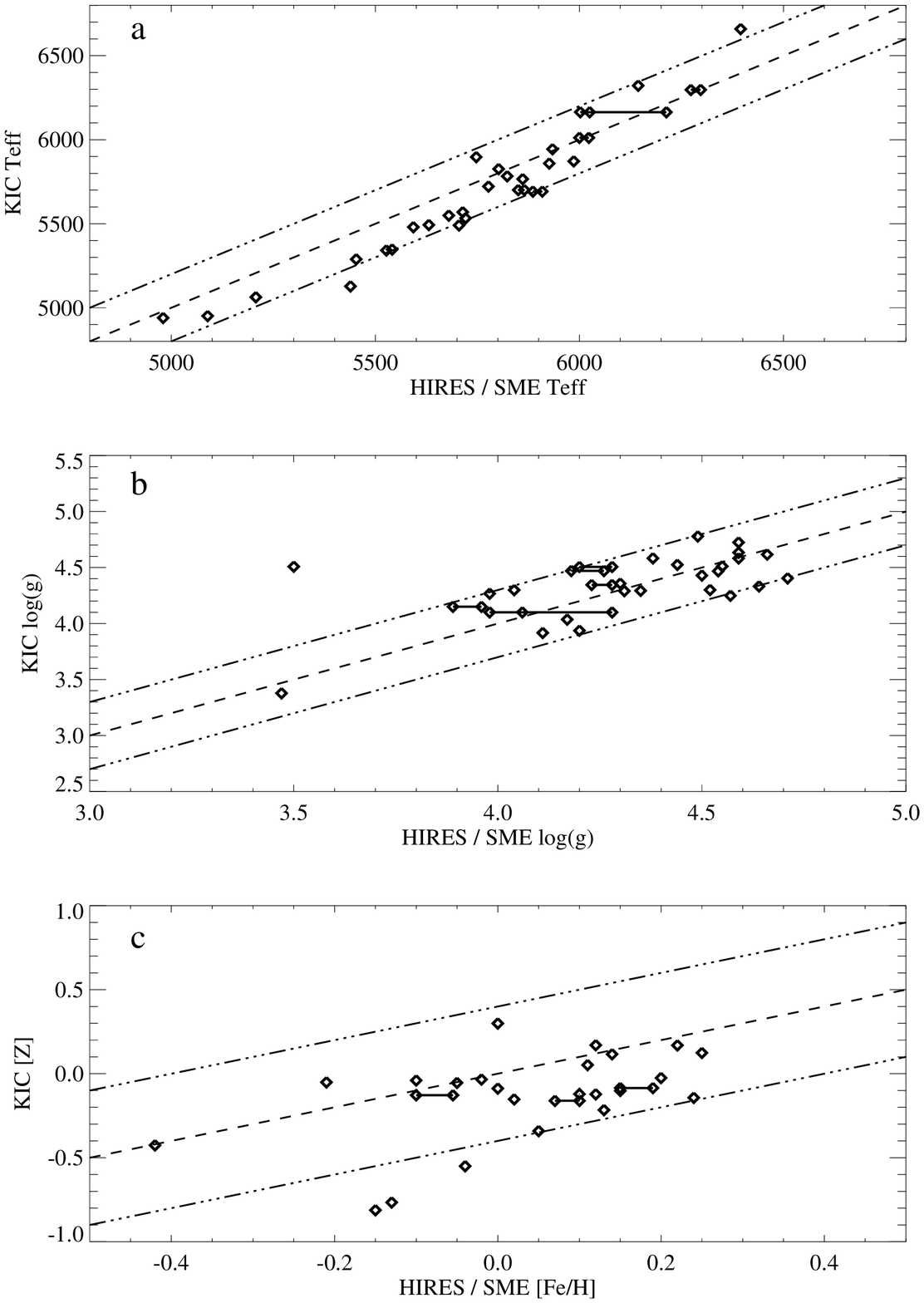}
\end{figure}

\clearpage

\begin{figure}
\plotone{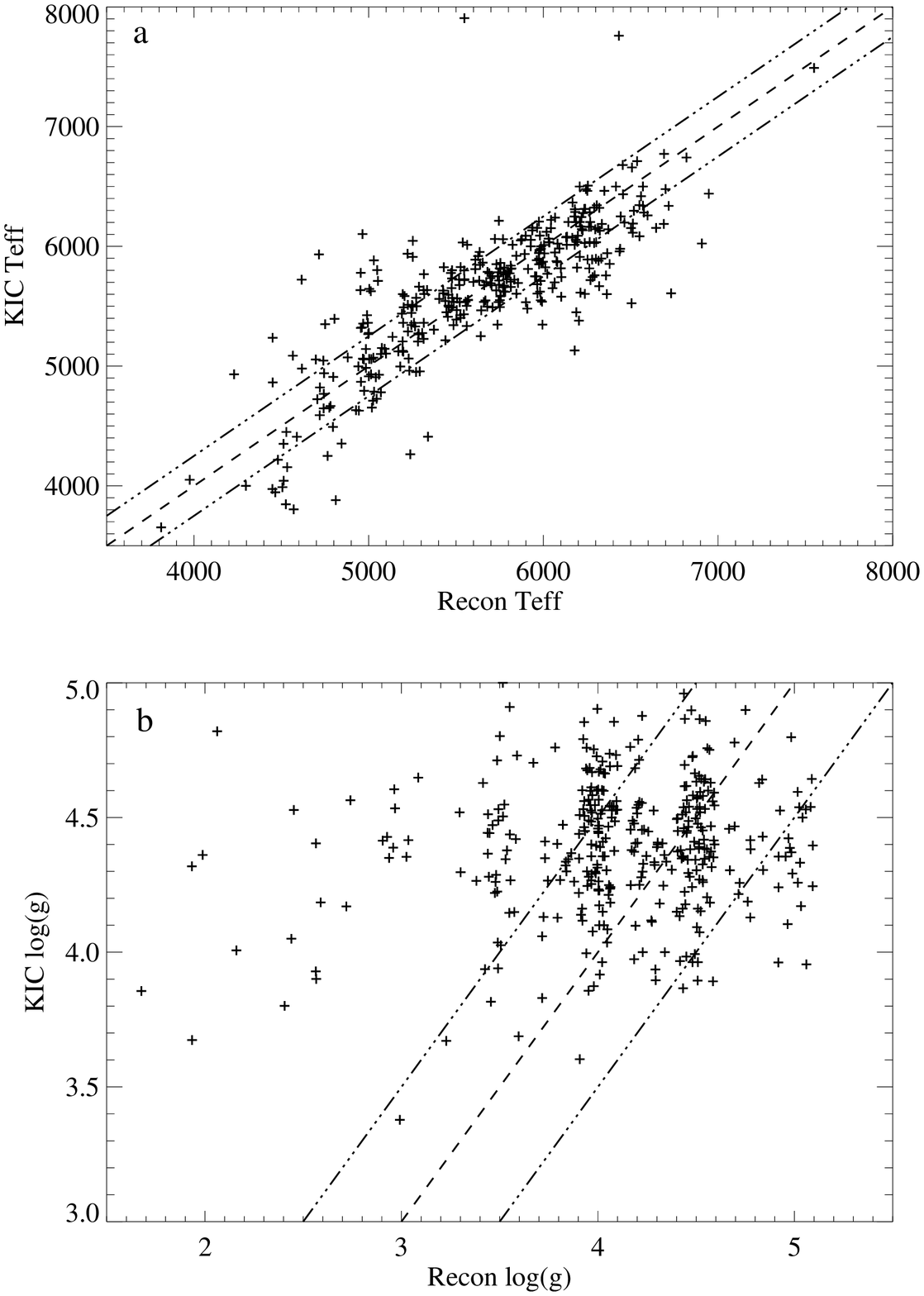}
\end{figure}

\clearpage

\begin{figure}
\plotone{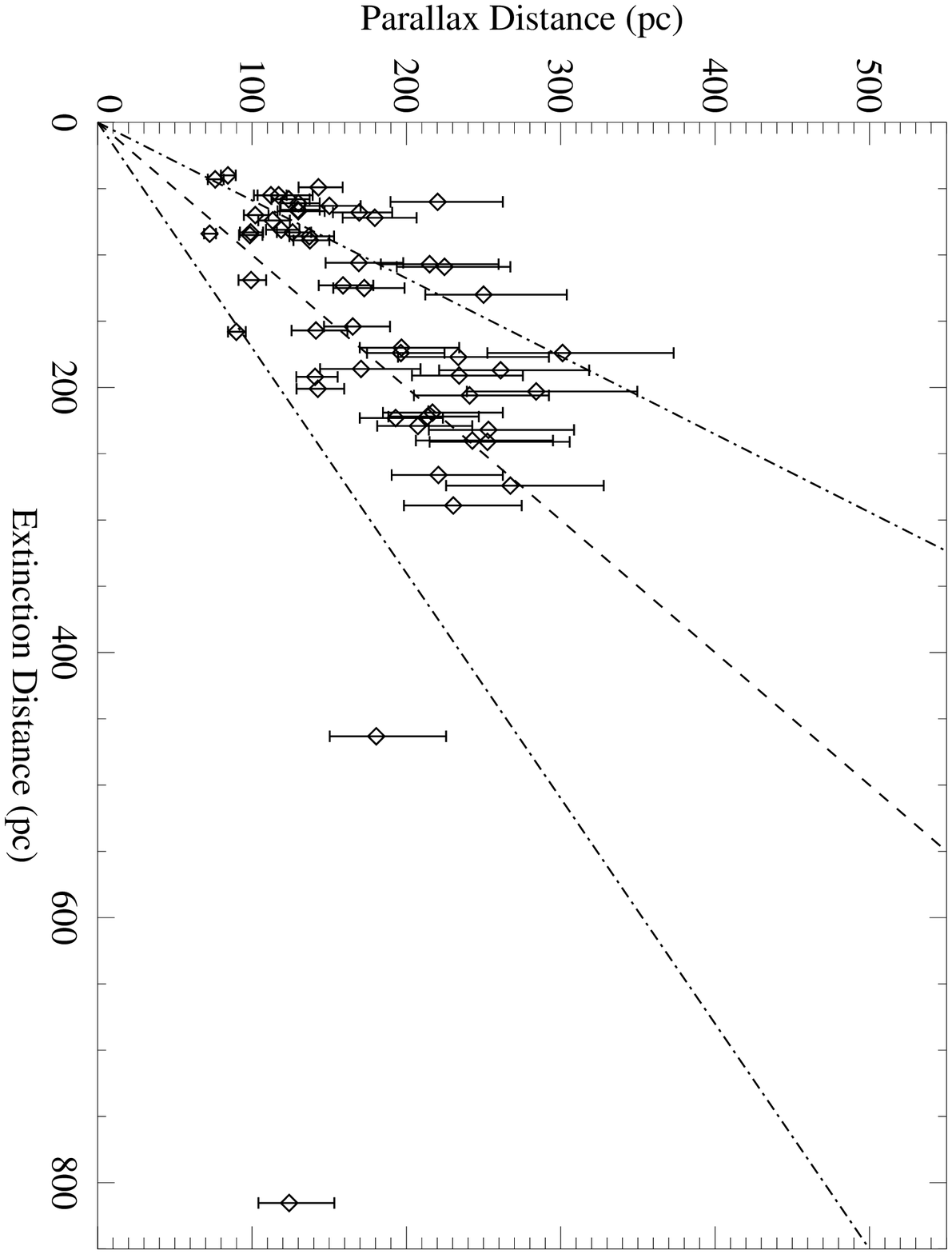}
\end{figure}

\clearpage

\begin{figure}
\plotone{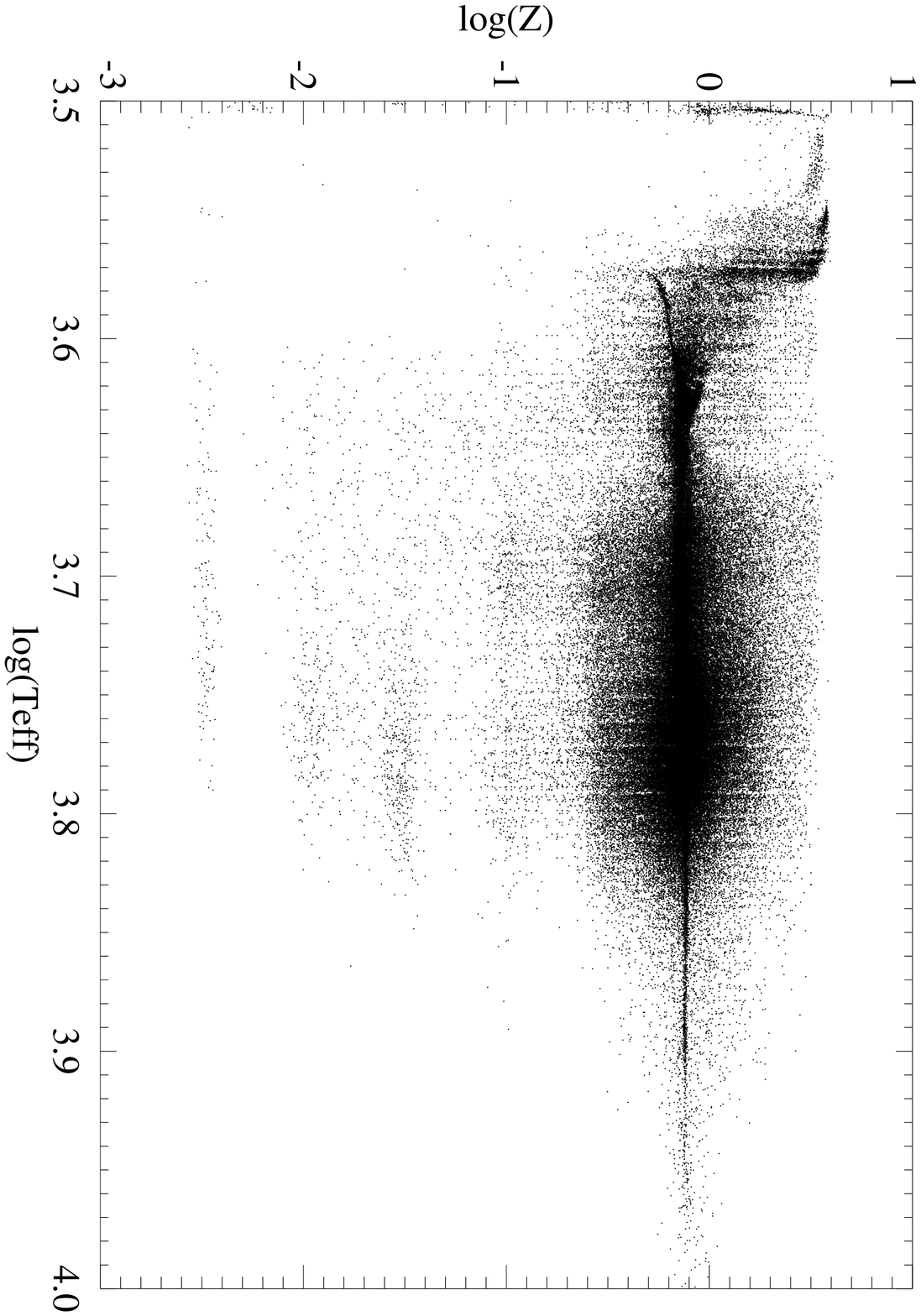}
\end{figure}

\begin{figure}
\plotone{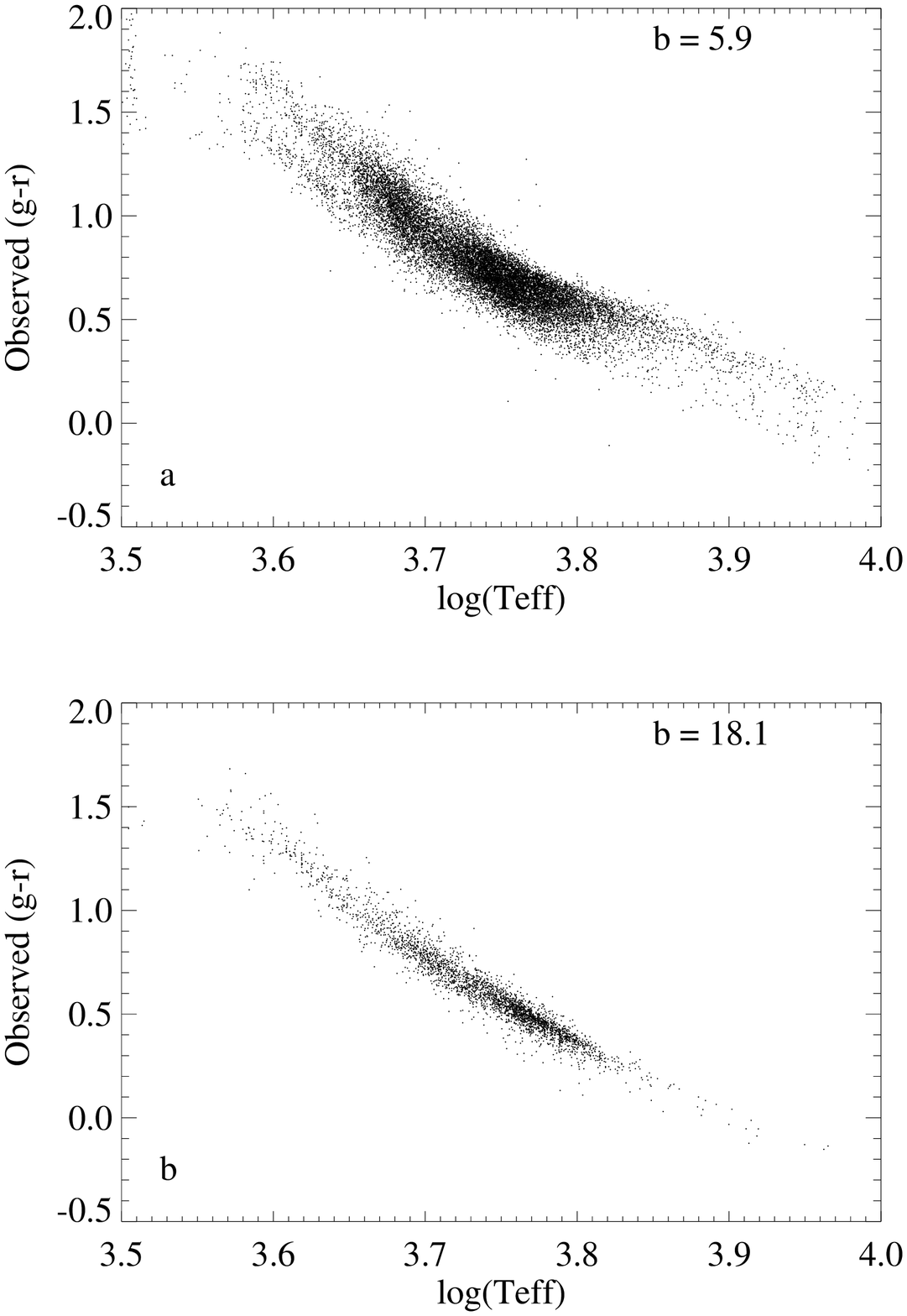}
\end{figure}

\clearpage

\begin{deluxetable}{rrrrrrrrrrrrrrrr}
\tabletypesize{\scriptsize}
\rotate
\tablecaption{KIC primary standard stars. \label{tbl-1}}
\tablewidth{600pt}
\tablehead{
\colhead{RA} & \colhead{Dec}   & \colhead{u}   &
\colhead{g} & \colhead{r} & \colhead{i} & \colhead{z} & \colhead{D51} &
\colhead{J} & \colhead{H} & \colhead{K} & \colhead{K$_{\rm{p}}$} & \colhead{$\teff$} &
\colhead{$\log(g)$}  & \colhead{$\log(Z)$} & \colhead{$\log(R_*)$} 
}
\startdata
 112.83372 &  29.07580 & 17.829 & 16.077 & 15.434 & 15.226 & 15.141 & 15.838 & 14.221 & 13.842 & 13.773 & 15.481 &  5386. &  4.483 & -2.540 & -0.021 \\
 112.83971 &  29.02154 & 16.745 & 15.344 & 16.282 & 14.527 & 14.407 & 15.519 & 13.493 & 13.037 & 12.996 & 16.047 &  6057. &  4.303 & -0.187 &  0.093 \\
 112.84009 &  29.07181 & 16.619 & 15.187 & 14.871 & 14.405 & 14.308 & 15.034 & 13.361 & 13.044 & 12.941 & 14.640 &  6023. &  4.334 & -1.577 &  0.076 \\
 112.85278 &  29.14279 & 17.426 & 16.191 & 15.712 & 15.549 & 15.502 & 15.995 & 14.657 & 14.362 & 14.285 & 15.742 &  6026. &  4.441 & -1.955 &  0.019 \\
 112.86972 &  28.85981 & 17.497 & 16.380 & 16.038 & 15.898 & 15.861 & 16.220 & 15.039 & 14.877 & 14.701 & 16.043 &  6257. &  4.404 & -1.608 &  0.043 \\
 112.87161 &  29.14310 & 18.962 & 17.127 & 16.438 & 16.205 & 16.144 & 16.876 & 15.185 & 14.740 & 14.739 & 16.482 &  5234. &  4.539 & -2.502 & -0.057 \\
 112.87298 &  28.92986 & 18.938 & 16.629 & 15.625 & 15.245 & 15.023 & 16.296 & 13.951 & 13.371 & 13.272 & 15.660 &  4614. &  4.583 & -1.682 & -0.129 \\
 112.87424 &  29.00252 & 17.530 & 16.366 & 15.968 & 15.828 & 15.772 & 16.191 & 15.005 & 14.787 & 14.676 & 15.990 &  6201. &  4.427 & -1.835 &  0.030 \\
 112.88787 &  29.02796 & 17.541 & 15.929 & 15.306 & 15.030 & 14.932 & 15.696 & 13.989 & 13.586 & 13.525 & 15.300 &  5559. &  4.426 & -2.504 &  0.017 \\
 112.88976 &  28.88614 & 17.904 & 16.576 & 16.065 & 15.903 & 15.816 & 16.372 & 14.951 & 14.608 & 14.494 & 16.105 &  5988. &  4.451 & -1.995 &  0.013 \\
\enddata
\end{deluxetable}

\end{document}